\documentclass[twocolumn]{aastex631}

\newcommand{\rtwo}{R$_{200}$}

\newcommand{\mtwo}{M$_{200}$}

\newcommand{\xhi}{$\chi_{HI}$}
\newcommand{\dndz}{\textit{d$\mathcal{N}$/dz}}

\newcommand{\msun}{$M_{\odot}$}

\newcommand{\rvir}{$r_{vir}$}

\newcommand{\hone}{H~\textsc{i}}

\newcommand{\cfour}{{\rm C}~\textsc{iv}}

\newcommand{\mgtwo}{Mg~\textsc{ii}}

\newcommand{\osix}{{\rm O}~\textsc{vi}}

\newcommand{\nhi}{n$_{HI}$}

\DeclareUnicodeCharacter{2212}{-}

\submitjournal{ApJ}

\shorttitle{A survey of H~\textsc{I} and O~\textsc{VI} absorption in galaxy cluster outskirts} 
\shortauthors{Holguin Luna et al.}
\graphicspath{{./}{figures/}}

\begin{document}

\title{A Survey of H~\textsc{I} and O~\textsc{VI} Absorption Lines in the Outskirts of $z\lesssim0.3$ Galaxy Clusters}

\correspondingauthor{Priscilla Holguin Luna}
\author{Priscilla Holguin Luna}
\affiliation{Department of Astronomy, New Mexico State University, Las Cruces, NM 88001, USA}
\email{pholg@nmsu.edu}

\author{Joseph N. Burchett}
\affiliation{Department of Astronomy, New Mexico State University, Las Cruces, NM 88001, USA}

\author{Daisuke Nagai}
\affiliation{Department of Physics, Yale University, New Haven, CT 06520, USA}
\affiliation{Department of Astronomy, Yale University, New Haven, CT 06520, USA}

\author{Todd M. Tripp}
\affiliation{Department of Astronomy, University of Massachusetts, Amherst, MA 01003, USA}

\author{Nicolas Tejos}
\affiliation{Instituto de F\'{i}sica, Pontificia Universidad Cat\'{o}lica de Valpara\'{i}so, Casilla 4059, Valpara\'{i}so, Chile}

\author{J. Xavier Prochaska}
\affiliation{Department of Astronomy 'I\&' Astrophysics, UCO/Lick Observatory, University of California, Santa Cruz, CA 95064, USA}
\affiliation{Kavli Institute for the Physics and Mathematics of the Universe (Kavli IPMU), 5-1-5 Kashiwanoha, Kashiwa, 277-8583, Japan}
\affiliation{Division of Science, National Astronomical Observatory of Japan, 2-21-1 Osawa, Mitaka, Tokyo 181-8588, Japan}
\affiliation{Simons Pivot Fellow}



\begin{abstract}

The intracluster medium (ICM) in the far outskirts (r $>$ 2--3 \rtwo) of galaxy clusters interfaces with the intergalactic medium (IGM) and is theorized to comprise diffuse, multiphase gas. This medium may hold vital clues to clusters' thermodynamic evolution and far-reaching impacts on infalling, future cluster galaxies. The diffuse outskirts of clusters are well-suited for quasar absorption line observations, capable of detecting gas to extremely low column densities. We analyze 18 QSO spectra observed with the Cosmic Origins Spectrograph aboard the \textit{Hubble Space Telescope} whose lines of sight trace the gaseous environments of 26 galaxy clusters from within \rtwo~to 6 \rtwo~in projection. We measure the \textit{dN/dz} and covering fraction of \hone~and \osix~associated with the foreground clusters as a function of normalized impact parameter. We find the \dndz~for \hone~is consistent with the IGM field value for all impact parameter bins, with an intriguing slight elevation between 2 and 3 \rtwo. The \dndz~for \osix~ is also consistent with the field value (within 3$\sigma$) for all impact parameter bins, with potential elevations in \dndz~both within 1--2 \rtwo~and beyond 4 \rtwo~at $>2\sigma$. We propose physical scenarios that may give rise to these tentative excesses, such as a buildup of neutral gas at the outer accretion shock front and a signature of the warm-hot IGM. We do not find a systematic excess of potentially associated galaxies near the sightlines where \osix~is detected; thus, the detected \osix\ does not have a clear circumgalactic origin.

\end{abstract}




\section{Introduction}\label{sec:intro}

In the nearby universe ($z <$ 0.1), the hot gas of the intracluster medium (ICM) is typically assumed to have reached thermal equilibrium within some virial radius. Work by \cite{sereno_2021} shows that galaxy clusters thermalized around a redshift $z \sim 0.14$, implying that higher redshift clusters are subject to physical processes, such as accretion and mergers, that prevent thermal relaxation from occurring at earlier times. The seminal work by \cite{butcher_1984} was the first to provide evidence that the environments of galaxy clusters vary with redshift, where it was found that blue galaxies are abundant in clusters at higher redshifts ($z \gtrsim 0.4$), yet are rare in the populations of nearby clusters ($z \lesssim 0.1$) that are dominated by red galaxies. \cite{dressler_1980} showed that the morphology of cluster members depends upon their location in the cluster, with the fraction of spirals decreasing in the dense core of the ICM as the fraction of S0 and ellipticals increases. It has since been shown that quenching, the cessation of star formation in a galaxy, can be seen in cluster members at high radii ($>$2--3 \rvir; \citealt{tonnesen_2007,bahe_2013,castignani_2022,lopes_2024}). The redshift evolution of star-forming cluster members, known morphology-density relation, and evidence of quenched satellites at large clustocentric radii implicate the cluster outskirts as a region of high interest, since the ICM in this region influences both cluster-scale evolution and the evolution of galaxies within the cluster. Through an in-depth study of this elusive environment, we can better understand the evolutionary pathways and mechanisms at play.

Observations of \hone~in the cluster environment reveal a dearth of \hone~relative to galaxies in less dense environments (Lyman-$\alpha$ absorption; \citealt{yoon_2012,yoon_2017,burchett_2018}), yet the absorption is detected in regions with nearby cluster members, or detected directly within the gaseous halos of cluster members (21-cm and Lyman-$\alpha$ absorption; \citealt{gavazzi_2005,yoon_2013,yoon_2017}), called the circumgalactic medium (CGM). The CGM plays a vital role in galactic ecosystems, as the gas that enters and exits a galaxy interacts directly with the halo. There is a known ``\hone~deficiency" in clusters, wherein cluster members show a dearth of cool \hone~in their interstellar medium (ISM) in comparison to galaxies in less dense environments. This has historically been shown using 21-cm observations for spirals in nearby clusters such as Virgo and Coma (\citealt{davies_1973,chamaraux_1980,sullivan_1981}). This \hone~deficiency aligns with the picture of a hot, virialized ICM, as evidenced by the tidal tails extending from ram-pressure stripped satellites in clusters, called jellyfish galaxies (\citealt{degrandi_2016,poggianti_2017,vulcani_2018,george_2022}). The deficiency is observed to have radial dependence, with satellites closer to the core containing less \hone~than those further out. In general, cluster members typically contain less \hone~than comparable field galaxies (\citealt{giovanelli_1983,magri_1988}). However, there is some tension regarding the presence of neutral hydrogen in these environments, as some 21-cm observations find excess \hone~in the regions surrounding clusters (\citealt{denes_2014}), yet both \hone~rich and \hone~deficient satellites have been observed in cluster outskirts (\citealt{chung_2009}). 

Ultraviolet (UV) quasar (QSO) absorption line studies that target \hone~Lyman-$\alpha$ depict a similar narrative for nearby clusters. This method uses the spectrum of a bright background QSO to study absorption signatures from the foreground gas. Because absorption is not as biased by gas density in the same way as emission, absorption is vastly more powerful than emission for probing low-density gas, such as that in the ICM. This method has been used to show that cluster members have a gas-deficient CGM compared to their counterparts in the field, and this deficiency increases with increasing environmental density (\citealt{yoon_2013,burchett_2016,burchett_2018}). Circumgalactic and intracluster Ly$\alpha$ absorbers are highly ionized, but the gas traced by Ly$\alpha$ absorbers can originate in cool (10$^4$ K) photoionizaed gas or warm-hot (10$^5-6$ K) collisionally ionized (e.g., shock-heated) environments. \cite{yoon_2012} \& \citealt{yoon_2017} found that Lyman-$\alpha$ absorbers detected around the Coma \& Virgo clusters, preferentially found beyond the cluster virial radii, were most likely associated with infalling gas from the cosmic web and had velocity dispersions indicative of turbulence in the outskirts of the ICM.

More recently, \cite{mishra_2024} reported the detection of Ly$\alpha$ in the outskirts of galaxy clusters and groups (M$_{halo} \geq 10^{13}$ \msun) from z = 0.01--0.76 through an analysis of stacked UV spectra. They too find a deficiency of cool, neutral gas in the CGM of cluster members closer to the cores of galaxy clusters, with the distribution and strength of Ly$\alpha$ absorbers indicating the presence of cool gas in the outskirts. Any identified, associated metal absorbers have very low covering fractions compared to the abundances detected in the CGM of field galaxies. These findings are in line with the picture painted by absorption studies of clusters at lower redshifts, with the implication of cool gas in the outskirts an intriguing motive for future studies. 

Studies of higher redshift ($z > 0.4$) clusters and luminous red galaxies, which can be viewed as proxies for massive group centrals, show an intriguing surplus of gas-rich CGM, often surrounded by large overdensities of gas (\citealt{overzier_2016,cai_2017,zahedy_2017,muzahid_2017,cai_2018,berg_2019,chen_2019,smailagic_2023}). Although this excess of \hone~Ly$\alpha$ may seem to be inconsistent with the known \hone~deficiency in groups and clusters discussed above, we note that LRG studies typically have a higher median redshift, so this may indicate an environmental transition or temporal evolution occurring between satellites and gas reservoirs in group- and cluster-mass halos. This transition would affect the ICM content between 0.5 $> z >$ 0, amplifying the importance of studying the ICM and galaxies in the outskirts throughout this epoch. 

Studies of the diffuse gas in galaxy clusters have involved a wide array of multi-wavelength observations, starting with observations of optical emission-line filaments in clusters (see \citealt{donahue_2022} for a review). More recently, observations of the diffuse ICM beyond \rtwo~in clusters has been revolutionized by  X-ray and Sunyaev-Zel'dovich (SZ) observations. The ICM detected via X-rays is frequently the near-virialized component, as the diffuse gas in the outskirts (r $>$ \rtwo) is difficult to detect in emission with current facilities due to the low X-ray surface brightness in this region. However, there have been time-intensive observations pushing X-ray detection to increasingly larger extents (\citealt{simionescu_2011,walker_2012,tchernin_2016,mernier_2017,eckert_2019,ettori_2019}, with \citealt{mernier_2018,walker_2019} for reviews). The gas measured by X-rays typically has temperatures from 10$^{6}$ -- 10$^8$ K (\citealt{cowie_1983,vikhlinin_2006,nagai_2007}). At the same time, simulations predict that the outskirts of the cluster contain both cool and warm-hot gas (\citealt{burns_2010,nagai_2011,emerick_2015,butsky_2019}), presenting a more complex, multiphase picture of the ICM. \cite{butsky_2019} finds that $>80\%$ of the gas mass at T = 10$^4$--10$^5$ K is beyond 2 \rtwo, implying that the outskirts of the clusters are rich with a component of the ICM extremely difficult to detect observationally due to its diffuse and relatively low density conditions (e.g., currently invisible in X-ray) and unfavorable temperature for X-ray studies. Whereas X-ray emission decreases proportionally to density squared, the SZ effect diminishes linearly with electron pressure, meaning that SZ observations have an advantage in detecting the hot gas further into the cluster outskirts over X-ray observations. The Atacama Cosmology Telescope (ACT) has been used to observe clusters beyond 1--2 \rtwo~(\citealt{sifon_2016}), with more recent ACT-SZ observations using a stacking analysis to observe the far outskirts ($>$3 \rtwo) of clusters (\citealt{hurier_2019,anbajagane_2022}). New SZ experiments, such as Simons Observatory and CMB-S4, will likely lead future detections of hot gas in cluster outskirts; however, the cool and warm-hot gas that dominates this regime is especially well-suited for UV quasar absorption line spectroscopy, which provides significantly higher spectral resolution (e.g., for kinematics) and information about the chemical composition of the cool gas.

In the far outskirts of clusters, the ICM meets the intergalactic medium (IGM), the gaseous component of the large-scale structure that forms the connective tissues of the cosmic web. This interface between the cool IGM and multiphase ICM leads to a non-negligible contribution of non-thermal pressure support (\citealt{evrard_1990, nagai_2007a, nelson_2014a}) to the ICM that is more significant in the outskirts of clusters (\citealt{lau_2009,battaglia_2012,avestruz_2016a,shi_2016a}), which are dominated by accretion processes that give rise to multiphase media (\citealt{nelson_2014,battaglia_2015}). The primary sources of non-thermal pressure support in cluster outskirts are gas motions, such as turbulence, that cause inhomogeneities in the density distribution of the ICM as the gas enters the ICM via filamentary streams (\citealt{burns_2010,zinger_2016}) or the CGM of infalling galaxies (\citealt{tonnesen_2009,bahe_2013,zinger_2018b}). The outermost accretion shock marks the boundary of a phase change between the hot ICM and the warm-hot intergalactic medium (WHIM) or the cool IGM. As infalling gas and galaxies must pass through this cluster region during accretion (though not all necessarily interact with the shock front), it is a potential source of density fluctuations in the ICM outskirts.

Unlike inner virial shocks that arise from the infall of galaxies and gas into the core of the cluster (\citealt{hurier_2019}), the outermost accretion shock is a remnant of the collapse of the primeval halo that formed the cluster. A hierarchical formation scenario in which density perturbations led to the gravitational collapse and heating of the cluster core (\citealt{peebles_1970,silk_1978}) is supported by the self-similarity predicted and observed in the entropy profile of clusters (\citealt{kaiser_1986,lau_2015,voit_2005} for a review). Although it is ubiquitously predicted in cluster formation simulations, the location of the shock varies from model to model, generally identified beyond the virial radius between 1 -- 5 \rtwo~(\citealt{molnar_2009,lau_2015,zinger_2018a,aung_2021,walker_2019}). Clusters are known to contain a higher quenched fraction of galaxies than in the field (\citealt{dressler_1980,donnari_2021a}), but different quenching processes likely dominate in different regions of the cluster environment. For example, ram pressure stripping (RPS) of the ISM likely dominates in the core (\citealt{gunn_1972,tonnesen_2007,zinger_2018a}). The outermost accretion shock may also contribute to the quenching of satellites as a boundary of phase change in the ICM, existing at the multiphase interface of the hot, inner region, and the outer region, which contains more cool and warm-hot gas. Constraining the shock front's location is necessary for comparing to theory and assessing its impact on the multiphase ICM and satellites that interact with it.

Our study aims to constrain the physical conditions of the multiphase ICM using quasar absorption line spectroscopy to map \hone~and associated metals from within \rtwo~to 6 \rtwo. \cite{spitzer_1956} first theorized the use of this method to study galaxy halos, which was further built on by \cite{bahcall_1969}, and decades of work using absorption line spectroscopy have revolutionized our understanding of the IGM (\citealt{tripp_2008,tejos_2016,danforth_2016,kim_2021}), the CGM (\citealt{tumlinson_2013,werk_2013,burchett_2018,prochaska_2019}), and the connection between them (\citealt{prochaska_2011,tejos_2014,burchett_2015,werk_2016}). Note that in this work, we adopt R$_{200c}$, the radius at which the density of the cluster is 200 times that of the critical density of the universe at the redshift of the cluster, as the virial radius. Here, we further explore this connection in the cluster environment using data from the Hubble Spectroscopic Legacy Archive to study 26 quasar-cluster pairs (Table \ref{tab:clusters}) in the local universe that have been observed by the UV-sensitive Cosmic Origins Spectrograph aboard the Hubble Space Telescope (HST/COS).

In Section \ref{sec:sampleinfo}, we discuss sample selection and data, followed by Section \ref{sec:dataanalysis}, which discusses data analysis and calculations. Section \ref{sec:results} presents our results, and we discuss the results in Section \ref{sec:discussion}. We summarize in Section \ref{sec:conclusion}. In this paper, we adopt the WMAP9 $\Lambda$CDM flat cosmology (\citealt{hinshaw_2013}), with $\Omega_M$ = 0.2855, $\Omega_\Lambda$ = 0.7135, and H$_0$ = 69.32 km s$^{-1}$/Mpc.

\begin{figure}
  \centering
  \includegraphics[width=0.5\textwidth]{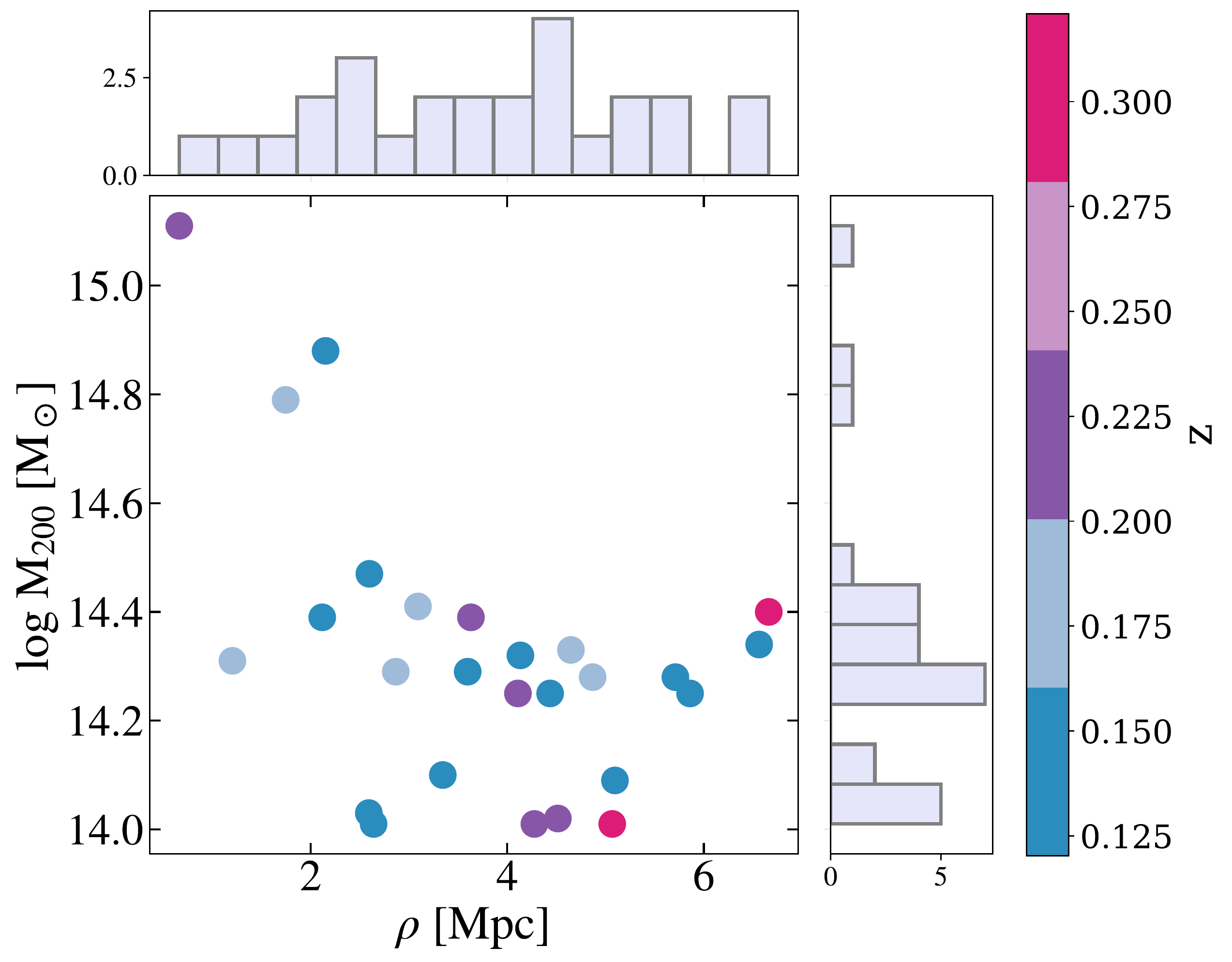}
  \caption{The mass distribution of the clusters in our sample plotted against their background QSO impact parameter ($\rho$). The points are color-coded by their redshift, with the lowest redshift cluster in our sample at 0.12 and the highest at 0.32. We see that the two highest redshift clusters are both probed beyond 6 Mpc from the cluster center, and there are few systems for which we probe within \rtwo. The range of \rtwo~spanned by our sample is $\sim$880--2000 kpc.}
  \label{fig:sample}
\end{figure}

\begin{table*}[t]
  \centering
  \caption{Table with information about each cluster-QSO system in our sample. Included is cluster name (from redMaPPer catalog), M$_{200c}$, R$_{200c}$, redshift, richness $\lambda$, and calculated velocity dispersion $\sigma_{disp}$, as well as the QSO redshift and system impact parameter in projected proper distance $\rho$, and normalized by cluster virial radius $\rho/R_{200c}$.} 
\begin{tabular}{ccccccccc}
\hline
Cluster ID & $z_{cluster}$ & log M$_{200c}$ & R$_{200c}$ & $\lambda$ & $\sigma_{disp}$ & $z_{QSO}$ & $\rho$ & $\rho/R_{200c}$ \\
\hline
 &  &  $\mathrm{M_\odot}$ & $\mathrm{kpc}$ & & $\mathrm{km~s^{-1}}$ & & $\mathrm{kpc}$ &  \\
 \hline
 \hline
30907 & 0.21 & 14.31 & 1156 & 20.8 & 871 & 0.48 & 4108 & 3.7 \\
10214 & 0.16 & 14.39 & 1247 & 24.66 & 918 & 0.24 & 4647 & 3.9 \\
46 & 0.21 & 15.15 & 2209 & 130.1 & 1665 & 0.39 & 663 & 0.3 \\
33547 & 0.14 & 14.09 & 1004 & 13.01 & 729 & 1.13 & 2591 & 2.7 \\
12649 & 0.31 & 14.46 & 1247 & 28.63 & 993 & 1.13 & 6660 & 5.6 \\
2691 & 0.13 & 14.53 & 1405 & 33.24 & 1013 & 0.27 & 2597 & 1.9 \\
8791 & 0.14 & 14.4 & 1267 & 25.19 & 921 & 0.43 & 6562 & 5.4 \\
77073 & 0.24 & 14.08 & 961 & 12.74 & 736 & 0.43 & 4514 & 5.0 \\
82403 & 0.22 & 14.07 & 962 & 12.47 & 728 & 0.43 & 4276 & 4.7 \\
24201 & 0.13 & 14.16 & 1061 & 15.1 & 767 & 0.4 & 3344 & 3.3 \\
843 & 0.19 & 14.84 & 1750 & 65.76 & 1303 & 0.4 & 1745 & 1.0 \\
11725 & 0.19 & 14.37 & 1216 & 23.63 & 909 & 0.22 & 1203 & 1.0 \\
4241 & 0.13 & 14.45 & 1320 & 28.03 & 955 & 0.14 & 2117 & 1.7 \\
163 & 0.14 & 14.93 & 1908 & 79.67 & 1382 & 0.41 & 2151 & 1.2 \\
11497 & 0.12 & 14.34 & 1219 & 22.17 & 877 & 0.16 & 5710 & 4.9 \\
47627 & 0.2 & 14.34 & 1188 & 22.17 & 889 & 0.24 & 4868 & 4.3 \\
9713 & 0.23 & 14.45 & 1275 & 28.03 & 971 & 0.24 & 3630 & 3.0 \\
37036 & 0.14 & 14.15 & 1049 & 14.79 & 763 & 0.33 & 5095 & 5.1 \\
6097 & 0.13 & 14.35 & 1225 & 22.65 & 885 & 0.33 & 3597 & 3.1 \\
337305 & 0.32 & 14.07 & 925 & 12.47 & 742 & 0.47 & 5068 & 5.7 \\
4411 & 0.19 & 14.47 & 1312 & 29.25 & 979 & 0.47 & 3091 & 2.5 \\
9432 & 0.17 & 14.35 & 1208 & 22.65 & 891 & 0.34 & 2866 & 2.5 \\
5859 & 0.15 & 14.38 & 1245 & 24.14 & 908 & 0.34 & 4133 & 3.5 \\
43551 & 0.16 & 14.07 & 982 & 12.47 & 720 & 0.32 & 2640 & 2.8 \\
7048 & 0.12 & 14.31 & 1191 & 20.8 & 858 & 0.19 & 5859 & 5.1 \\
8183 & 0.12 & 14.31 & 1193 & 20.8 & 857 & 0.47 & 4436 & 3.9 \\
\hline
\end{tabular}
  \label{tab:clusters}
\end{table*}

\section{Sample Description} \label{sec:sampleinfo}
To identify galaxy cluster-QSO pairs, we imposed the following criteria: our analysis requires spectral coverage of \hone~Ly$\beta$ (1025 \AA) to corroborate Ly$\alpha$ identification, as well as coverage of the \osix~doublet (1032 \& 1038 \AA) which traces 10$^{5.5-6}$ K warm-hot gas. In cases where Ly$\alpha$ is saturated, we adopt a lower limit for N(\hone) (see Section \ref{subsec:linemeas}). HST/COS provides moderate-resolution yet highly sensitive spectra in the UV wavelength range 1150--3200 \AA. The minimum redshift of clusters in our sample is $z\sim$0.1 to cover the lines above, and we have imposed an upper redshift limit of $z\sim$0.3 to maximize sample size while minimizing evolutionary timescales. Therefore, the cluster redshifts range from 0.1--0.3, with the median redshift of the sample at $z_{clust}\sim$0.18. Our quasars fall in the redshift range $z_{QSO}$ = 0.14--1.2, with the median redshift of the quasars at $z_{QSO}\sim$0.4. Key information about each QSO-cluster system is provided in Table \ref{tab:clusters}.

The velocity difference between the QSO-cluster system is at minimum $\Delta$v = 5000 km s$^{-1}$ to ensure identified absorbers are beyond the velocity range of excess absorption potentially associated with the QSO (\citealt{tripp_2008}). Galaxy cluster velocity dispersion can range from 100--1500 km s$^{-1}$ (\citealt{bahcall_1981,struble_1991}), with a typical cluster velocity dispersion of $\geq750$ km s$^{-1}$ within $\sim$1--2 Mpc (\citealt{bahcall_1996}). As our study extends to 6 \rtwo~($\simeq$5.2 and 12.8 Mpc for the smallest and largest \rtwo~in our sample respectively), to capture infalling material, we consider the upper end of this velocity dispersion range and consider identified lines within 1500 km s$^{-1}$ of the velocity of the cluster to be kinematically associated with the cluster. Figure \ref{fig:sample} shows the mass distribution of the cluster sample as a function of the impact parameter and is color-coded by the cluster redshift. To mitigate potential biases that could arise from knowledge of cluster locations during the line identification and fitting process, there was no prior knowledge of potential foreground cluster redshifts during line identifications in the QSO spectra.

Our sample leverages archival data from the Hubble Spectroscopic Legacy Archive (HSLA) and consists of 18 sightlines that probe 1--3 foreground clusters each within 6 \rtwo, providing data for 26 clusters. Our 26 UV bright QSO-cluster pairs were identified by cross-matching galaxy clusters with M$_{200}$ $>$ 10$^{14}$ M$_\odot$ (M$_{200}$ derived from optical richness $\lambda$; \citealt{rykoff_2012}) from the redMaPPer catalog (\citealt{rykoff_2014}) with existing QSO spectra in the HSLA.  We use the redMaPPer catalog reported $\lambda$ and $z$ values to calculate each cluster's systemic velocity, \rtwo, and M$_{200}$ -- the latter two calculated as described in \cite{rykoff_2012} -- and refer the reader to \cite{rykoff_2014} for a detailed description of the algorithms used to determine optical richness and $z$. The QSOs in our sample were therefore not selected for cluster observations, but blindly selected for other sciences cases. Table \ref{tab:clusters} contains relevant cluster and QSO parameters for each system in our sample. The lower-$z$ clusters in the sample have sufficient spectral coverage in G130M grating observations, but we require observations in the G160M grating to cover \hone~Ly$\alpha$ for higher z clusters. We require that each spectrum used for analysis have S/N $\gtrsim$ 10 in order for the equivalent width limit of Ly$\alpha$ to be W$_r$ = 50 m\AA, which corresponds to the expected \hone~column density, N(\hone), of weak absorbers in these regions: 10$^{13}$ cm$^{-2}$ (\citealt{butsky_2019}). This sets the minimum N(\hone) for our measurement of \textit{dN/dz} -- the redshift number density of absorbers, a means of statistically measuring absorption incidence of a given strength per unit redshift, and that is further discussed in Section \ref{subsec:dndzcalc}.

\section{Analysis} \label{sec:dataanalysis}

\subsection{Line Identification} \label{subsec:lineid}
Before identifying absorption lines in our QSO spectra, we fit a continuum to each spectrum using the \textsc{linetools} package\footnote{https://github.com/linetools/linetools} (\citealt{prochaska_2016}). Once the spectrum was normalized, we identified the absorption components using the \textsc{igmguesses} graphical user interface within the \textsc{pyigm} package\footnote{https://github.com/pyigm/pyigm}, which allows users to identify the absorption lines in velocity space and apply a rough fit with Gaussian profiles. The methodology for identifying lines in a spectrum was adapted from \cite{tejos_2016}, with some minor adjustments, described as follows. We assigned each identified line to one of three categories: (a) reliable, (b) possible, and (c) uncertain, according to the following procedure:
\begin{enumerate}
\itemsep0em 
    \item Identify all commonly identified Milky Way lines in HST/COS bandpass at $z$ = 0 within $\pm$ 250 km s$^{-1}$ and mark them as reliable.
    \item Identify all \hone~and metal lines within $\pm$ 250 km s$^{-1}$ of the redshift of the quasar. Mark \hone~lines with at least two transitions (e.g., Ly$\alpha$ and Ly$\beta$, or Ly$\beta$ and Ly$\gamma$) and associated metal lines aligned in velocity with \hone~as reliable.
    \item Working from $z$ = $z_{QSO}$ backward to $z$ = 0, identify all \hone~absorption components with at least two transitions present and mark them as reliable.
    \item Identify metal lines within $\pm$ 250 km s$^{-1}$ of each identified \hone~absorption component and mark as reliable. When multiple transitions are covered for a single ion, the relative positions of the components must be consistent in velocity space and have similar kinematic structure to be marked as reliable. For cases in which blending may be present, we use the strength of a non-blended absorption component for an ion to determine the expected relative line strength\footnote{For a given preliminary fit to a line in \textsc{igmguesses}, a profile for other lines present for that same ion are shown based on the fit of that first line.} for the potentially blended line and mark this as possible if there are multiple other non-blended lines present for that ion or uncertain if there is only one other non-blended line for that ion. We return to potentially blended lines after all absorbers in the spectrum have been identified to make any necessary adjustments to the reliability category given further line identification.
    \item We determine whether the remaining unidentified absorption features could be \hone~Ly$\alpha$ with associated metal lines, and mark these as as probable. All other unidentified absorbers are marked as \hone~Ly$\alpha$ with an uncertain ranking.
\end{enumerate}

The identification of lines is ranked based on whether there is a corroborating line at the same redshift. For example, if a potential Ly$\alpha$ line is found in sufficient redshift, there should be a corresponding Ly$\beta$, Ly$\gamma$, Ly$\delta$, etc., with consistent velocity and strength. The same requirement of line corroboration is true for identified metal ions to be marked as reliable. Still, we did not identify metal absorbers unless there is a- or b-ranked \hone~Lyman absorption identified at that redshift. While we acknowledge that metal absorbers can be reliably identified without an accompanying \hone~Lyman series detection, we do not search for metals unassociated with \hone~in our analysis. A possible rank identification is assigned to absorbers that do not have a corroborating line, either because of a lack of wavelength coverage or potential blending. An uncertain rank is used to identify a species that does not have a corroborating line, because the corroborating line is too weak to detect, likely due to insufficient S/N.

After identifying all potential absorption lines and systems, the line list was run through a 1D clustering algorithm that groups absorption systems within 750 km s$^{-1}$ of each other. Any identified absorption within 1500 km s$^{-1}$ of the systemic velocity of the galaxy cluster is considered kinematically associated with it. Upon completion of the identification process, the redshifts of the foreground clusters are incorporated into the analysis for the first time and solely as a reference for determining the kinematic association between the cluster and the identified absorption systems. 

We will describe absorption systems as follows: Multiple absorption components may belong to an identified system of absorbers. An absorption system can contain multiple components of multiple species. For example, there may be three Ly$\alpha$ detections for a single system --- this means there are three absorption components identified within the absorption system, each will contain at least a Ly$\alpha$ line but may also contain other covered Lyman lines at that redshift and identified associated metal absorbers. Therefore, within each component, there may be multiple lines. As we do not search for metals without \hone, our identified absorbers will always contain at least one \hone~Lyman line and may contain some metals.

\begin{figure}
  \centering
      \includegraphics[width=0.5\textwidth]{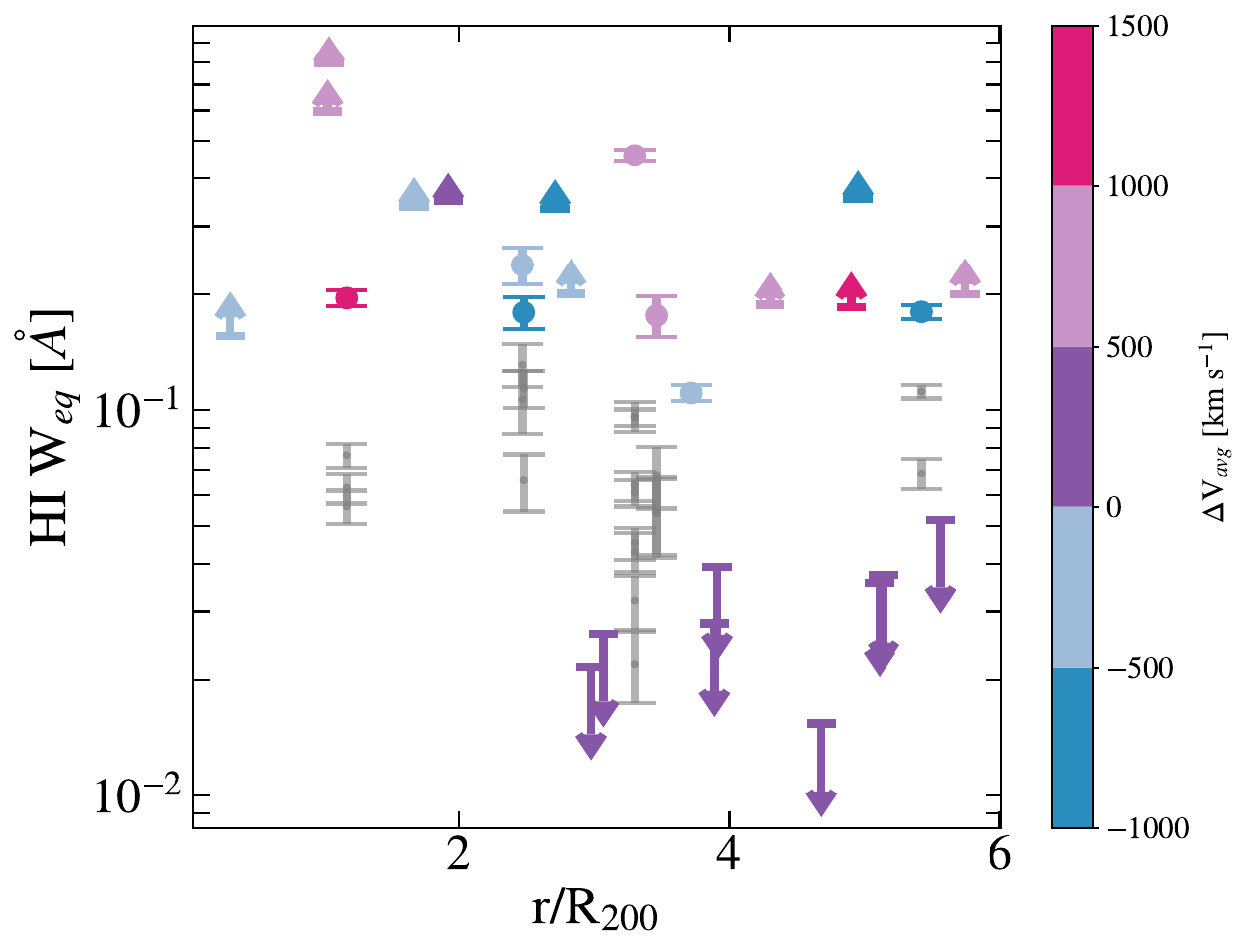}
  \caption{The W$_r$ of \hone~Ly$\alpha$ absorbers associated with clusters in our sample as a function of their impact parameters normalized by \rtwo. Detections are marked by circles, with upward pointing triangles for lower limits, and downward pointing triangles for upper limits. For systems with more than one component of Ly$\alpha$ $\lambda$1216 \AA~ identified, the depicted W$_r$ corresponds to the sum of the equivalent widths for all statistically significant components, with the light grey markers beneath depicting the individual absorption components that make up the sum. The point's color corresponds to the velocity difference between the identified absorber(s) and the host cluster, with the average velocity used for multi-component systems. The detections with the strongest cumulative W$_r$, which are not saturated lower limits, are found within 4 \rtwo, with upper limits only found beyond $\sim$3 \rtwo. }
  \label{fig:HI_EW}
\end{figure}

\begin{figure}
  \centering
  \includegraphics[width=0.5\textwidth]{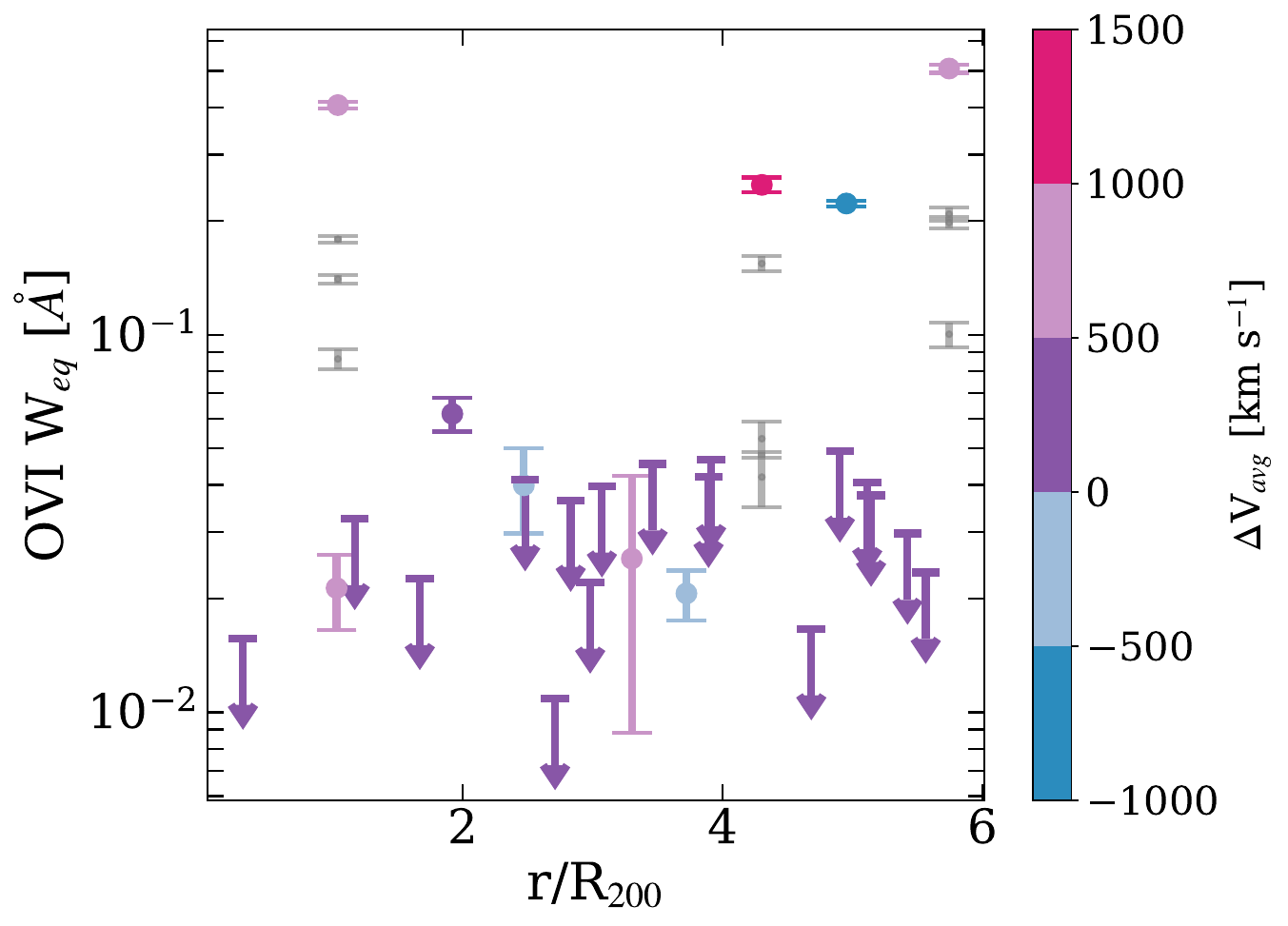}
  \caption{Similar to Figure 3 for \osix~$\lambda$1032 absorbers. Like the W$_r$ for \hone, multi-component absorption systems reflect the summed equivalent widths and the color reflects the average velocity separation. For summed systems, the light grey markers beneath the colored point depict the individual absorption components that make up the sum. Triangles mark upper limits, and detections are shown by circles. Most clusters do not exhibit \osix\ absorption, as shown by the fairly even spread of upper limits across the impact parameter range covered by our study. }
  \label{fig:OVI_EW}
\end{figure}

\subsection{Line measurements}
\label{subsec:linemeas}
We use \textsc{linetools} to perform equivalent width and apparent optical depth measurements (\citealt{savage_1991}) in an automated fashion, the latter being used for column density calculations. These are performed on lines identified as associated with the cluster-QSO system and within 1500 km s$^{-1}$ of the cluster's systemic velocity. For non-detections, we instantiate an absorption component using \textsc{linetools} at the redshift of the cluster. These components are treated as upper limits with $\pm$50 km s$^{-1}$ velocity limits. We use the measured rest-frame equivalent width, W$_r$, to determine line saturation. For Ly$\alpha$ lines with W$_r >$ 0.143 \AA, a lower limit N(\hone) = 10$^{13.75}$ cm$^{-2}$ is adopted. This equivalent width and corresponding column density were determined by convolving 1000 idealized, simulated Ly$\alpha$ absorbers with the COS line spread function and examining the spread in known N(\hone) for a measured equivalent width. The chosen W$_r$ saturation threshold corresponds with a 0.3 dex spread in column density at log (N(\hone)/cm$^{-2}$) = 13.75.

There are 8 clusters in our sample where we did not detect \hone~within 1500 km s$^{-1}$ from the cluster's redshift. The summed equivalent width (W$_{eq}$) measurements for identified \hone~Ly$\alpha$ absorbers in a single QSO-cluster system is shown in Figure \ref{fig:HI_EW}, where they are color-coded by the absorber's velocity relative to the systemic velocity of the cluster. Similarly, Figure \ref{fig:OVI_EW} depicts the summed equivalent width measurements for \osix\ on a QSO-cluster system level. There are 17 clusters in our sample in which we do not detect \osix~absorption within 1500 km s$^{-1}$ of the cluster's systemic velocity. All of the fundamental measurement information for the identified absorbers in our sample are listed in Table \ref{tab:absorbers}, found in the Appendix.

\subsection{\dndz~Calculation and Analysis} \label{subsec:dndzcalc}

The \textit{dz} for a sightline is defined as the total redshift path length over which it is possible to detect an absorption line of a given strength, i.e., equivalent width. The \textit{dz} for a spectrum is calculated by first converting the spectrum into redshift space as though there was an absorption feature for a given line at every wavelength. For example, to determine the redshift range over which we can detect a \hone~$\lambda1216$ absorber, we convert the wavelength at a pixel to $z$ assuming that the line falls at each pixel location, so $z = \lambda_{obs}/1215.67 - 1$. The regions in the spectrum where \hone~$\lambda1216$ is detectable are summed cumulatively, so the final \textit{dz} for a spectrum is the sum of all regions in the spectrum where \hone~$\lambda1216$ is detectable. The difference in redshift between each pixel is calculated. Limits are then placed on the minimum equivalent width detectable in groups of pixels based on the local signal-to-noise ratio, the limiting equivalent width, $W_{lim}$. This is calculated as follows:
\begin{equation}
    W_{lim}(\lambda) = \frac{3\sigma_{W(\lambda)}}{1 + z_{abs}} 
    \label{eqn:wlim}
\end{equation}
where $z_{abs}$ is the redshift corresponding to the wavelength and $\sigma_{W(\lambda)}$ is the uncertainty of the observed equivalent width, W$_\lambda$, summed in quadrature. We sum this over each pixel $i$:
\begin{equation}
    \sigma^2_{W(\lambda)} = \sum_i \left( \Delta \lambda(i) \left[ \frac{\sigma_{I(\lambda_i)}}{I(\lambda_i)} \right]  \right)^2
\end{equation}
where $\Delta \lambda_i$ is the pixel width in \AA~ and $I(\lambda_i)$ is the flux of the continuum at pixel $i$. Therefore, the $W_{lim}$ values will vary with pixel ranges and regions in the spectrum. 

\begin{figure}
  \centering
  \includegraphics[width=0.5\textwidth]{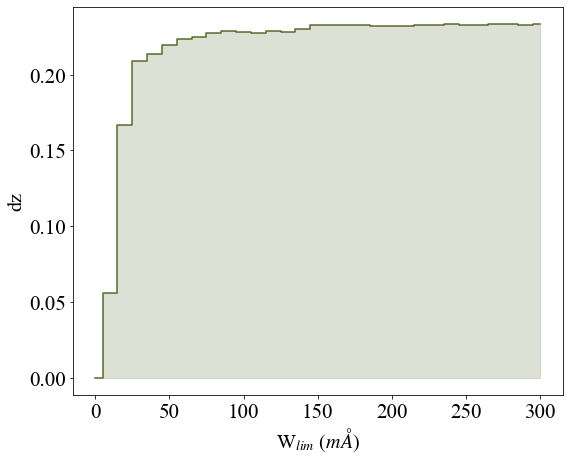}
  \caption{Path length, $\Delta z$, around the clusters in our sample as a function of limiting equivalent width, W$_{lim}$. The dark green line shows the total dz for all the clusters in our sample, with a knee appearing around 50 m$\AA$. This knee occurs as the path length reaches a maximum in dz, meaning that the W$_{lim}$ of the line is so strong beyond this it could be detected at just about all wavelengths in all of the spectra. The calculation is described in Section \ref{subsec:dndzcalc}, with the total \textit{dz} of the sample determined to be 0.215.}
  \label{fig:totaldz}
\end{figure}

Because absorption lines are detected over a range of pixels, rather than individual pixels, the number of pixels used to evaluate our chosen limiting equivalent width thresholds correspond to a range of pixels. In this work, 20--100 m\AA\ corresponds to 3--23 pixels, the range we use for analysis. Spectral regions with limiting equivalent widths less than a chosen threshold value (i.e., 20, 50, or 100 m\AA~for this study) are included in the total \textit{dz}. The total \textit{dz} for our sample is 0.215, shown in Figure \ref{fig:totaldz}.

The \dndz~is the number of statistically significant absorbers detected for a given species above some specified absorber strength (\textit{d$\mathcal{N}$}) within the redshift range over which we could detect that absorption (\textit{dz}). To find the \dndz~values for either \hone~or \osix, the rest equivalent width (W$_r$) uncertainty $\sigma_{W(\lambda)}$, redshift path length $\Delta z$, limiting rest equivalent width $W_{lim}$, and number of absorbers with EW greater than $W_{lim}$, \textit{d$\mathcal{N}$}, are necessary. The equation is as follows:
\begin{equation}
    \frac{d\mathcal{N}}{dz}(W) = \frac{\sum_i}{\sum_i} \frac{\mathcal{N}_i(W)}{\Delta z_i (W)}
\end{equation}
where $\mathcal{N}_i(W)$ is the number of absorbers in the \textit{i}-th equivalent width bin, $\Delta z_i (W)$ is the redshift path length for detecting absorbers, and all detections for which $W > W_{lim}$ are included in the summation.

We also calculate \textit{d$\mathcal{N}$/dz}$_{field}$ for multiple W$_{lim}$. This value is the \dndz~of the IGM for a given absorption strength measured on a number of random sightlines. Knowing \textit{d$\mathcal{N}$/dz}$_{field}$ allows for comparison between \dndz~for a given species and the ambient amount of absorption expected in the field for a given species. The equation we used to calculate was adapted from \cite{danforth_2016}:
\begin{equation}
    \frac{d\mathcal{N}(>N)}{dz} = C_{14} \left( \frac{N}{10^{14} cm^{-2}} \right)^{-(\beta - 1)}
    \label{eqn:dndz_field}
\end{equation}
where $N$ is the column density of \hone, $\beta$ = 1.65 $\pm$ 0.02 is the best-fit value to the differential distribution of absorbers in their study with 12 $\leq$ log $\left( \frac{N(\hone)}{cm^{-2}}\right) \leq$ 17, and C$_{14}$ = 25 $\pm$ 1 is the normalization constant for a fiducial column density N(\hone) $\sim$ 10$^{14}$ cm$^{-2}$. For \textit{d$\mathcal{N}$/dz}$_{field}$ of \osix, we use a normalization constant of C = 9.7 $\pm$ 1.3 and $\beta$ = 1.525 $\pm$ 0.26, the latter of which is the average $\beta$ for the two best-fit $\beta$ found by \cite{danforth_2016}, and correspond to a column density of 10$^{13.9}$ cm$^{-2}$. We use the limiting equivalent widths (20, 50, and 100 m\AA) to calculate (see Equations 9.14 and 9.15 in \citealt{draine_2011}) the approximate column density used in Equation \ref{eqn:dndz_field}. This approximation is accurate to within 2.6\% for $\tau_0 < 1.254$, where $\tau_0$ is the optical depth at line-center. Since all W$_{lim}$ values for which we use this approximation return column densities below our saturated column density threshold of $N_{sat} = 10^{13.75}$ cm$^{-2}$, the assumption that $\tau_0 << 1$, and we are therefore on the linear portion of the curve of growth, means the column density approximated with W$_{lim}$ is independent of a Doppler $b$ and $\gamma$.

\subsection{Covering fraction}
We calculate the covering fraction for \hone~and \osix~as a function of the projected impact parameter to the cluster's center, where we bin the data by virial radius-normalized impact parameter r/R$_{200}$ in bin widths of 2 \rtwo. If all clusters in a given bin have at least one statistically significant detection (i.e., above some W$_r$ threshold), the bin would have a covering fraction of 1. We adopt a W$_r$ threshold of 50 m\AA~for covering fraction calculations due to the saturation of dz that occurs at this value for our sample (see Figure \ref{fig:totaldz}). The uncertainties in covering fraction are estimated using the binomial confidence interval for each bin.

\begin{figure*}
  \centering
  \includegraphics[width=\textwidth]{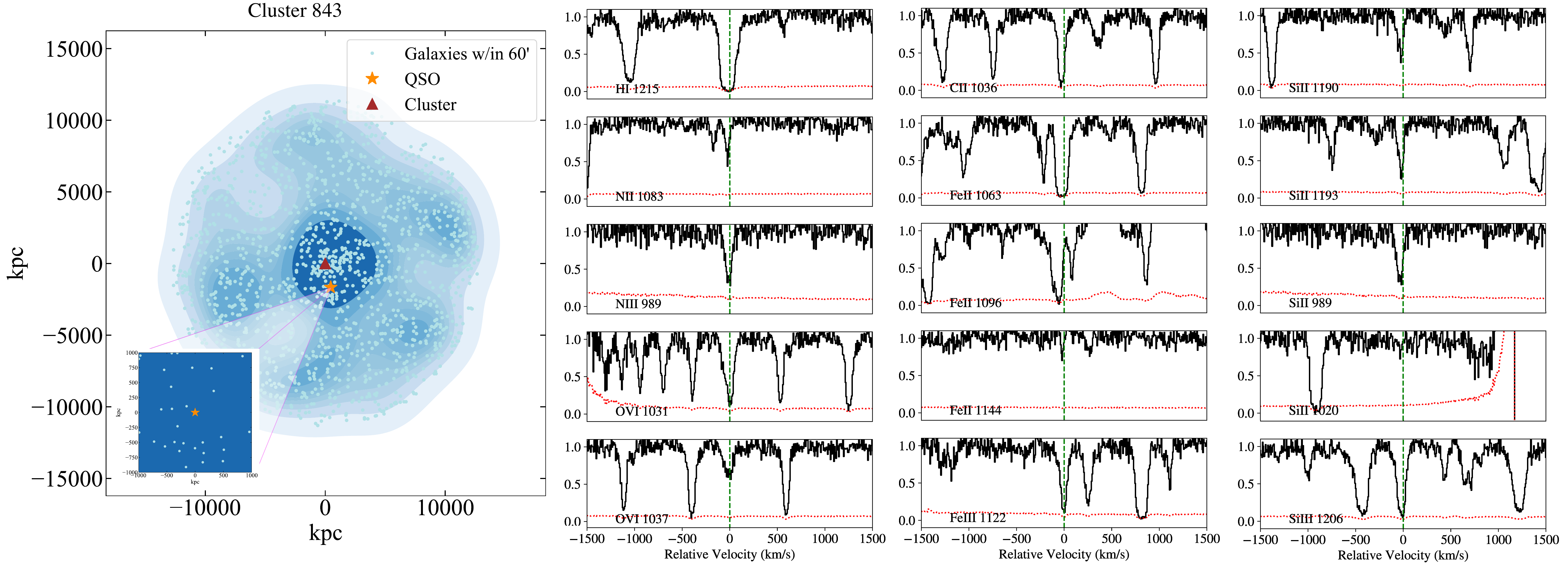}
\caption{\textit{Left}: The 2D galaxy density surrounding one of our metal-rich cluster-quasar systems. The inset of the left panel represents a zoomed-in 1000$\times$1000 kpc$^2$ distribution of galaxies around QSO. \textit{Right}: All identified absorbers in this system plotted with reference to the systemic velocity of the cluster.}
  \label{fig:metalrich}
\end{figure*}

\subsection{Local galaxy density}
We examine the identified metal absorbers in the context of the impact parameter and the density of cluster satellites in the ICM region probed by the QSOs. For metal-rich sightlines, the QSO may be probing not just the foreground ICM but also the CGM of nearby satellites, thus giving rise to a higher incidence of metal absorbers. To determine the density of cluster members surrounding the impact parameter at which a QSO pierces the ICM of the foreground cluster (and the potential CGM of a nearby cluster satellite), we queried SDSS for galaxies with and without spectroscopic redshifts projected within 6 \rtwo~of each cluster core. For those with spectroscopic redshifts (spec-zs), we used the SDSS photometric redshift (photo-z) assuming the galaxies were at the redshift of the cluster. Because our clusters are at redshifts where the SDSS spectroscopic completeness is low, we adopted the following method for identifying galaxies to be considered at a similar redshift to the cluster (see Table \ref{tab:clusters}):
\begin{enumerate}
    \item We calculated the velocity dispersion of the cluster using \mtwo~ and \rtwo~to identify the redshift range corresponding to each cluster and compared to all galaxies within 6 projected \rtwo~for which we had spectroscopic redshifts. In this distribution, we identified those within 3 times the velocity dispersion.
    \item Once we determined the spec-z range for cluster members, we identified a corresponding photometric-z (photo-z) range by down-selecting the spec-z sample only to include galaxies within the velocity dispersion of the cluster. This leaves on the order of 10 spec-z values in the down-selected sample. The corresponding photo-z range for this down-selected spec-z sample was adopted as the preliminary photo-z range for a cluster, keeping those with SDSS \textit{r-band} Petrosian magnitudes brighter than a magnitude limit of $r <$ 23.
    \item We then applied a cut to narrow the preliminary photo-z range using the standard deviation of the photo-z distribution, removing galaxies with photo-z deviating by more than 2$\sigma$ from the median \textit{spec-z} value. The remaining galaxies photo-zs composed the final photo-z range for that system. 
    \item The final photo-z range was applied to the full sample of galaxies for a given cluster-QSO system with $r <$23 to include both galaxies with and without measured spec-z values.
\end{enumerate}
To summarize, using the SDSS spectroscopic redshifts, we statistically determined a photometric redshift range that encompasses the uncertainties in photo-zs for a spectroscopic redshift range that falls within the velocity dispersion of the cluster. While we do our best to mitigate the inclusion of non-cluster member galaxies by implementing a magnitude limit and determining our photo-z range from spectroscopically confirmed cluster members, we acknowledge some likely contamination from outliers and interlopers in this sample. The left-most panel of Figure \ref{fig:metalrich} depicts an example of the galaxy density surrounding a cluster in our sample, with the identified absorbers in the system identified as associated with this cluster included on the right panels.

\section{Results} \label{sec:results}

We examine the dependences of \hone~and \osix~absorbers in covering fraction, \dndz, and equivalent width on the clustocentric impact parameter and galaxy density in the field.

\begin{figure}
  \centering
  \includegraphics[width=0.5\textwidth]{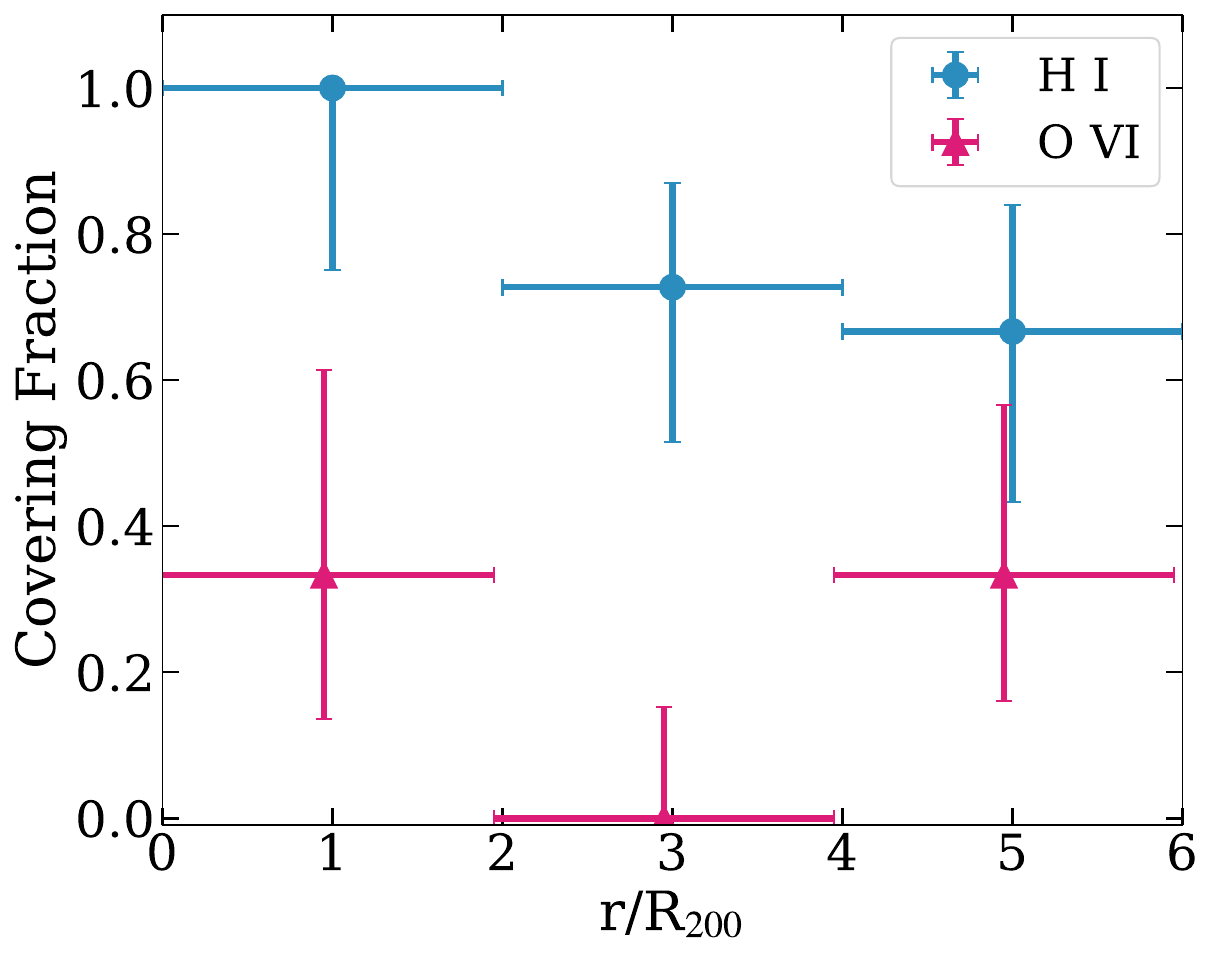}
  \caption{The covering fraction, using a chosen detection threshold of 50 m\AA, for \hone~and \osix. The error bar estimates represent the binomial confidence interval for each bin. We use three bins here (0--2, 2--4, and 4--6), to maximize number of objects per bin in our covering fraction calculation. For all bins $<$4 \rtwo, \hone~has a higher covering fraction than \osix. \hone~shows a value of unity within 2 \rtwo, while the covering fraction in the 2--4 \rtwo~bin is not unity due to the chosen threshold value. \hone~and \osix~have consistent covering fractions in the highest bin (4--6 \rtwo). \osix~shows non-zero covering fraction at 0--2 \& 4--6 \rtwo, which is consistent with the location of the slightly elevated \dndz~we detect for this ion.}
  \label{fig:cov_frac}
\end{figure}

\begin{figure*}
  \centering
  \includegraphics[width=\textwidth]{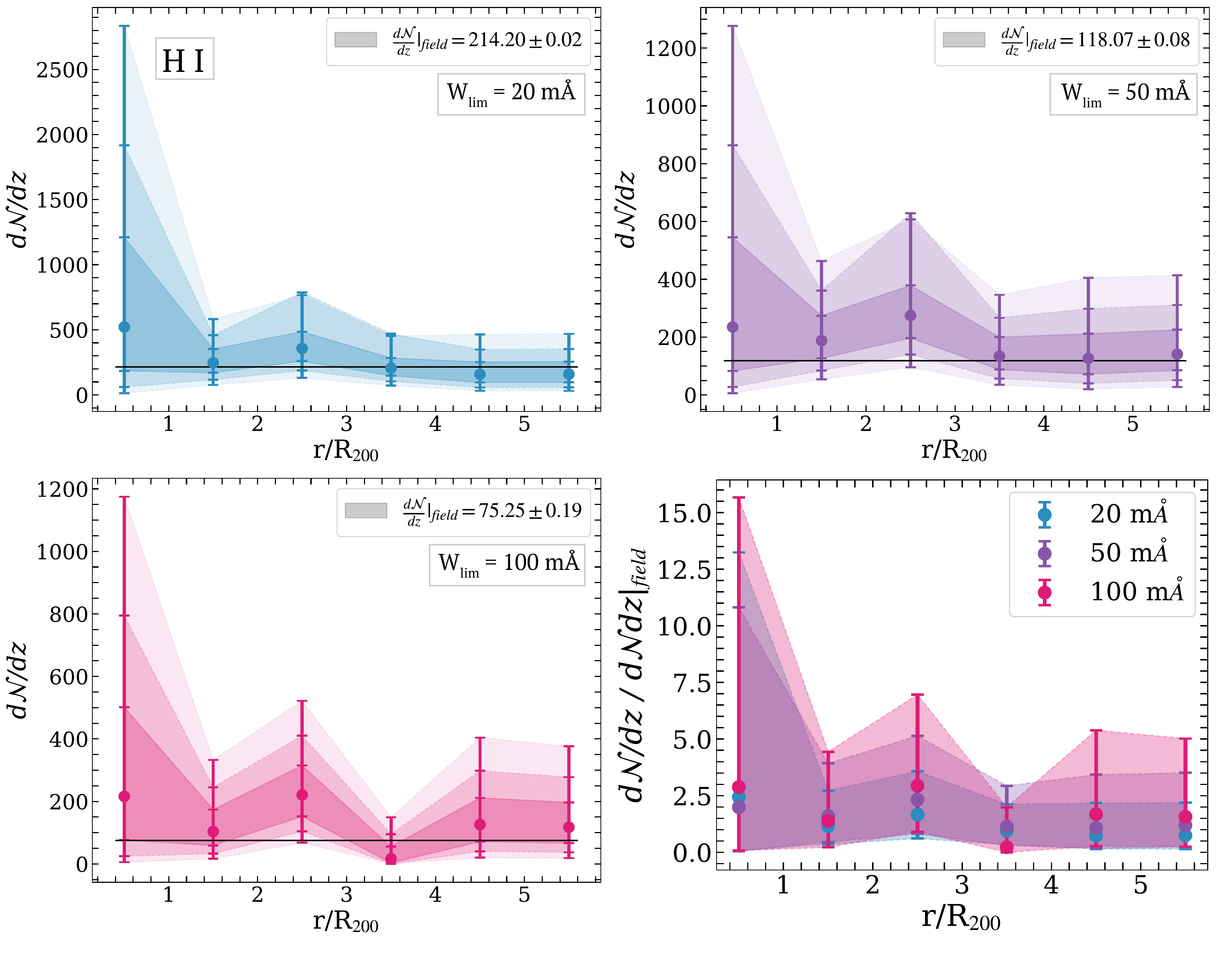}
  \caption{\textit{Upper \& lower left, upper right}: Dots mark the \dndz~values corresponding to each bin of r/\rtwo~at the W$_{lim}$ labeled in the upper right corner of each panel with the 1-, 2-, and 3$\sigma$ confidence intervals depicted per point by the shaded regions and errorbars. The black lines mark the expected \dndz$|_{field}$ value for absorption lines of the respective W$_{lim}$ strength, also listed in the upper right of each panel. Note the consistency in the \hone~profile shape for each figure, with the slightly ($>$ 2$\sigma$) elevated \dndz~above the field value between 2--3 \rtwo, which is most pronounced in the 100 m\AA~ figure. \textit{Lower right}: The normalized \dndz, with respective 3$\sigma$, for all W$_{lim}$ values.}
  \label{fig:HI_dndz}
\end{figure*}

\begin{figure*}
  \centering
  \includegraphics[width=\textwidth]{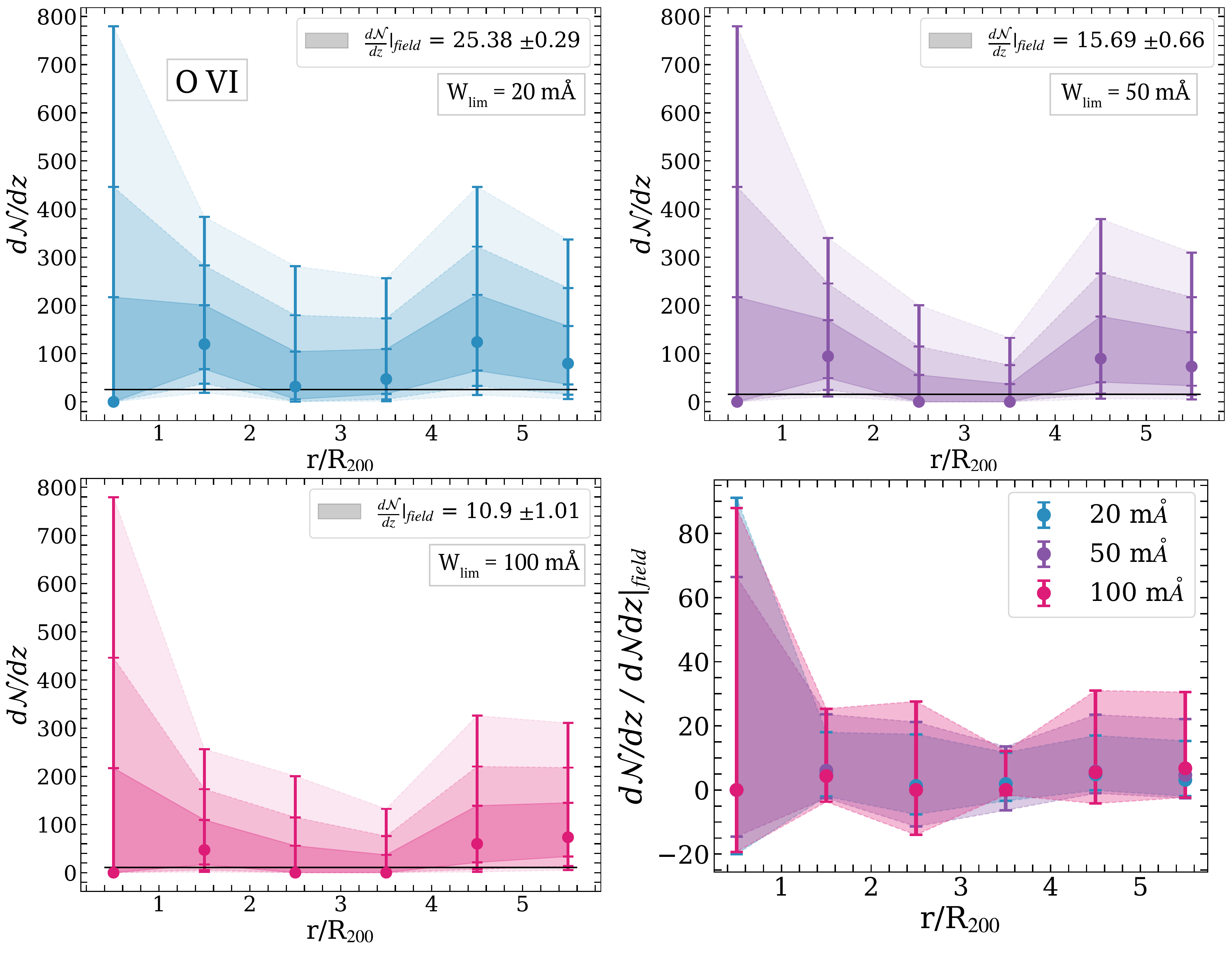}
\caption{\textit{Upper \& lower left, upper right}: Same as Figure \ref{fig:HI_dndz}, but for \osix. The \dndz~for \osix~is broadly consistent with the field value at every W$_{lim}$, with a slight ($>$ 2$\sigma$) elevation at 1--2 and 4--5 \rtwo~for W$_{lim}$ = 20 \& 50 m\AA. \textit{Lower right}: The \dndz~for all W$_{lim}$ measured normalized by \dndz$|_{field}$. The shaded region reflects the respective 3$\sigma$ confidence level for each W$_{lim}$.}
  \label{fig:OVI_dndz}
\end{figure*}

\subsection{Covering Fraction}
In Figure \ref{fig:cov_frac}, we see that within 4 \rtwo, \hone~has a higher covering fraction than \osix. Beyond 4 \rtwo, the covering fraction is consistent for \hone~and \osix~within their respective 1$\sigma$ errors, which are estimated using the binomial confidence interval for each bin. The equivalent width threshold for both \hone~and \osix~is 50 m\AA, and we use larger bins for the covering fraction analysis to alleviate uncertainties and bolster the statistics of each bin for comparison. Although, at first glance, the sub-unity covering fraction of \hone~from 2--4 \rtwo~may appear inconsistent with the location of the tentative \hone~excess in \dndz, we want to draw attention to Figure \ref{fig:HI_EW}, showing W$_{eq}$ as a function of \rtwo. Here we see that while the 2--4 \rtwo~bins contain many strong absorbers, it also contains two upper limits around 3 \rtwo, both of which fall below the 50 m\AA~ covering fraction threshold selected. These sub-50 m\AA~ upper limits bolster the denominator in the covering fraction calculation for the 2--4 \rtwo~bin, which drives the value below unity. There is thus a near-unity covering fraction for \hone~from 1--4 \rtwo, with all non-detections found in the outskirts beyond $\sim$3 \rtwo. 

The covering fraction of \osix~more closely follows the trend in \osix~\dndz, with detections within 2 \rtwo~and beyond 4 \rtwo. The strong absorbers detected in these bins, as seen in Figure \ref{fig:OVI_EW}, boost the values of the covering fraction. The trend in \dndz~for \osix, which reflects a slight elevation from 1--2 \rtwo~and beyond 4 \rtwo, is thus consistent with the trend seen in the covering fraction of \osix.

\subsection{\dndz}
Figure \ref{fig:HI_dndz} shows the \dndz~of identified \hone~absorbers at three different limiting equivalent widths. \dndz~was calculated following the process described in Section \ref{subsec:dndzcalc}, with the \dndz~value for each impact parameter bin represented by a dot and the accompanying 1, 2, and 3$\sigma$ confidence levels for each detection bin shown by the shaded errorbar regions. The $d\mathcal{N}/dz|_{field}$ value for each corresponding W$_{lim}$ is calculated using Equation \ref{eqn:dndz_field} and is marked by the black line in each figure. As a column density is required for the $d\mathcal{N}/dz|_{field}$ value to be estimated, we use an approximated column density for a given W$_{eq}$ (20, 50, and 100 m\AA), as described in Section \ref{subsec:dndzcalc}. At our lowest W$_{lim}$, 20 m\AA, we see an agreement within 1$\sigma$ between \dndz~and $d\mathcal{N}/dz|_{field}$, with the exception of the 2--3 \rtwo~bin showing a $>$1$\sigma$ variation between the amount of \hone~absorption detected and the amount expected from ambient \hone~absorption in the field. This increases to a $>$2 and 3$\sigma$ difference in \dndz~and $d\mathcal{N}/dz|_{field}$ for 50 and 100 m\AA~respectively. The strength of the detection increasing as W$_{lim}$ increases indicates that this tentative excess is arising from strong absorbers, though the statistical significance of this intriguing potential excess is not high and therefore requires confirmation.

The \dndz~of \osix, shown in Figure \ref{fig:OVI_dndz}, reveals fairly consistent values with $d\mathcal{N}/dz|_{field}$ across all absorber strengths. The black line in each figure shows the respective $d\mathcal{N}/dz|_{field}$ value for each corresponding W$_{lim}$. We detect a slight elevation in \dndz~for \osix~between 1--2 and 4--5 \rtwo~that is stronger for lower W$_{lim}$ ($>$2$\sigma$ for 20 \& 50 m\AA~and $>$1$\sigma$ for 100 m\AA). We see this suggestive excess for \osix~decreasing as W$_{lim}$ increases, suggesting the slight elevation in \dndz~is from weaker absorbers. Generally, the \dndz~of \osix~is consistent with the $d\mathcal{N}/dz|_{field}$ for all impact parameter bins, with an intriguing elevation detected to a 2$\sigma$ confidence for W$_{lim}$ = 20 \& 50 m\AA~at 1--2 \& 4--5 \rtwo.

\subsection{Trends in equivalent width}
We detect 49 statistically significant \hone~absorbers within 26 clusters. The frequency of absorbers detected per system increases between 2--4 \rtwo, with a higher number of weaker absorbers found per system, in these bins, and the highest non-saturated W$_{eq}$ value found in 3--4 \rtwo. This higher frequency of absorbers between 2--4 \rtwo\ is shown by the gray markers in this region in Figure \ref{fig:HI_EW}.

We detect 14 \osix~absorbers within 26 clusters, where all \osix~detections are associated with an \hone~Ly$\alpha$ detection. As stated in Section \ref{subsec:lineid}, we do not identify \osix\ absorbers independent of \hone, so all identified \osix\ is part of an absorption system that contains \hone\ absorption as well. The upper limits for \osix\ are spread across all \rtwo~bins, with a lack of statistically significant detections found from 2--4 \rtwo~(see Figure \ref{fig:OVI_EW}).

\subsection{Galaxy density in field}
We examined the galaxy density for each system in conjunction with the spectra showing all identified absorbers in that system. Figure \ref{fig:metalrich} depicts this, showing the galaxy field density and absorbers identified for a metal-rich system in our sample. We consider systems with metals other than \osix~identified as metal-rich. We find that for $\sim64\%$ clusters, the highest density region within 8 Mpc and 1500 km s$^{-1}$ of the cluster is consistent with the center of the cluster. For $\sim25\%$ of the systems, we find that the overdensity identified does not perfectly align with the core of the cluster, which could be due to either i) an overdensity not associated with the cluster but within the photo-z range is contaminating the data with a significant amount of non-cluster member galaxies, or ii) the cluster center coordinates from the redMaPPer catalog may not be accurate. Of the 5 metal-rich systems in our sample, we see that while the QSO may not always coincide with the densest region of the foreground cluster environment, the space between the cluster core and the impact parameter at which the QSO probes it contains a high density of galaxies. Thus, there appear to be ample sources from which the metal-rich absorbers we identify in these sightlines may originate, with some galaxies within a few hundred kpc from the projected impact parameter of the QSO. Upon visual inspection, we do not find any correlation between the number of metals identified in an absorption system and the galaxy density.

\begin{figure}%
    \centering
{{\includegraphics[width=0.5\textwidth]{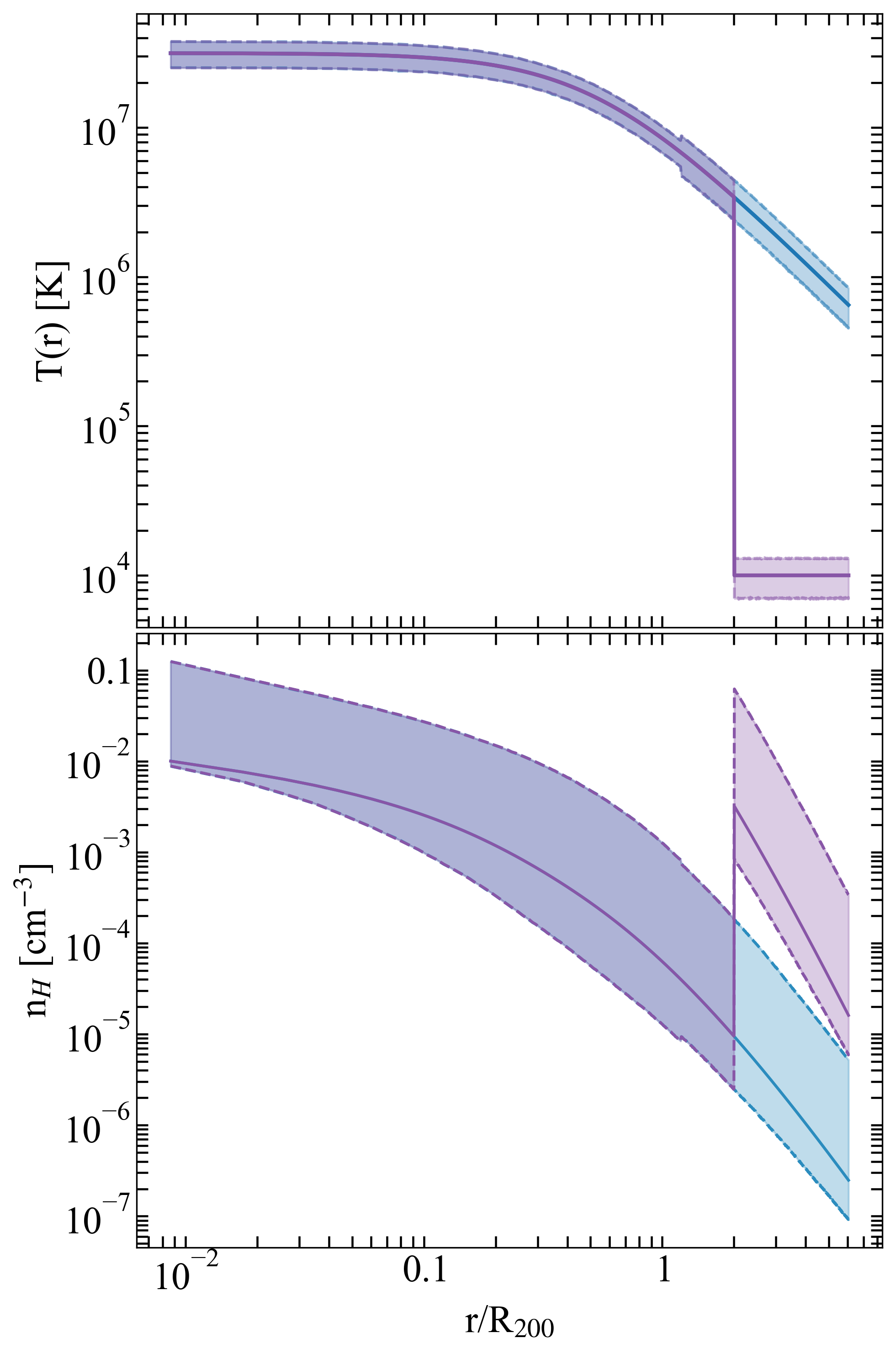} }}%
    \caption{\textit{Top:} The initial and modified temperature profiles used in our toy model. The solid blue line depicts the temperature profile adopted from \citealt{lau_2015}, which asymptotically approaches the temperature of the WHIM at larger r. The dashed purple line shows our modified temperature profile, where a discontinuity is introduced at r $\geq$ 2 \rtwo, forcing the gas temperature beyond this radius to be consistent with that of the IGM (10$^4$ K). \textit{Bottom:} The number density of hydrogen calculated using the \cite{lau_2015} temperature profile (solid blue line) and our modified temperature profile (dashed purple line). At 2 \rtwo, there is a density enhancement when the modified temperature profile is used. The lower temperature threshold of the modified profile allows for the subsistence of denser gas traced by \hone.  }%
    \label{fig:tempdens}%
\end{figure}

\begin{figure}
  \centering
  \includegraphics[width=0.5\textwidth]{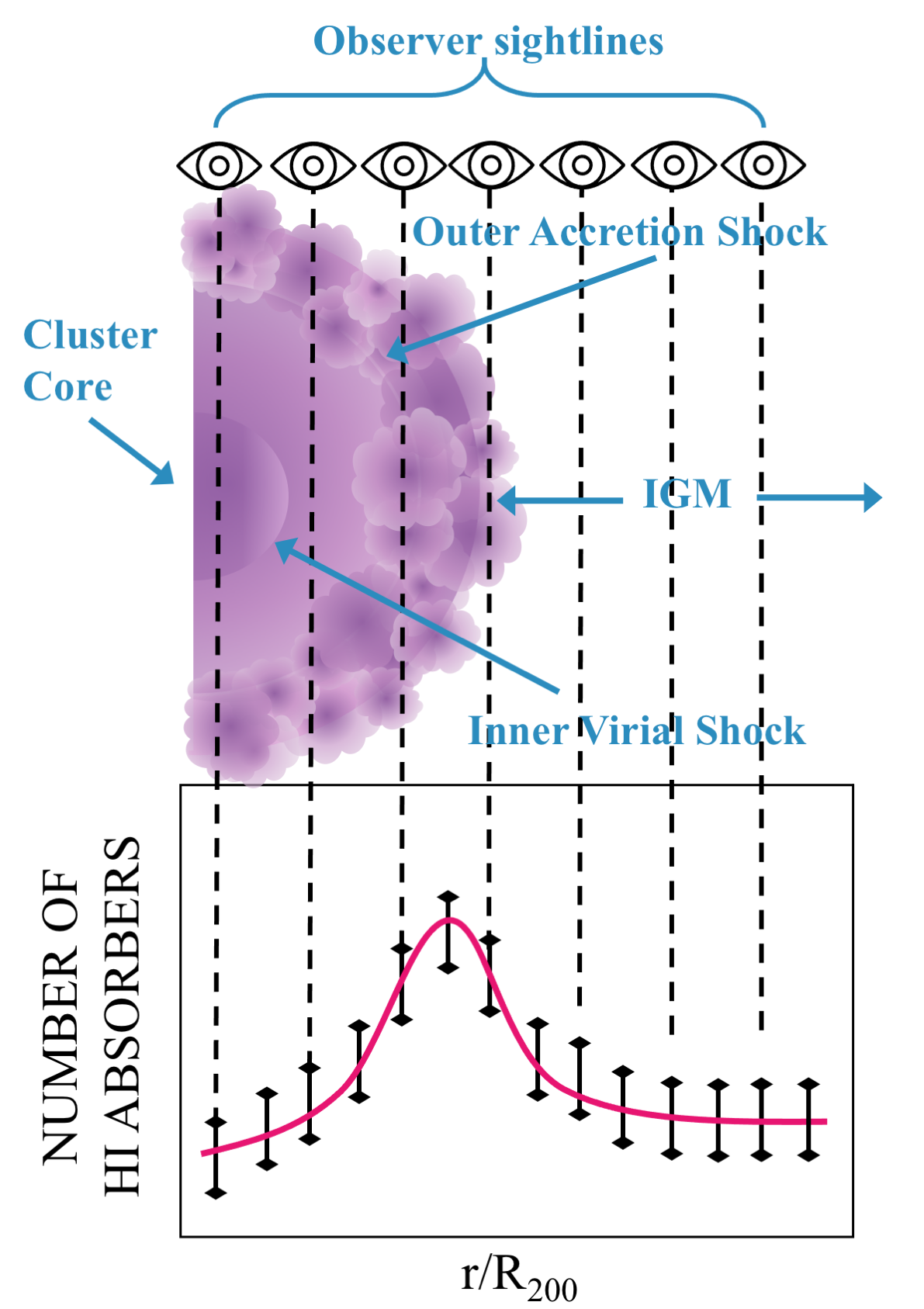}
  \caption{A schematic depicting the physical scenario in which a buildup of \hone~occurs at the accretion shock front. Note that scale and symmetry are dramatized for clarity. We expect to detect a higher amount of \hone~in sightlines that pierce the limb of the shock front, where they will encounter a higher amount of diffuse neutral hydrogen in the form of the CGM or IGM, represented by different sized `cloudlets' in the figure above. This diagram is an idealized scenario for a single cluster probed by multiple sightlines to depict the absorption trend (lower plot) one might expect. Therefore, this does not reflect the statistical nature of our analysis in which only 1--2 sightlines probe a single cluster.  }
  \label{fig:diagram}
\end{figure}

\begin{figure}%
    \centering
{{\includegraphics[width=0.5\textwidth]{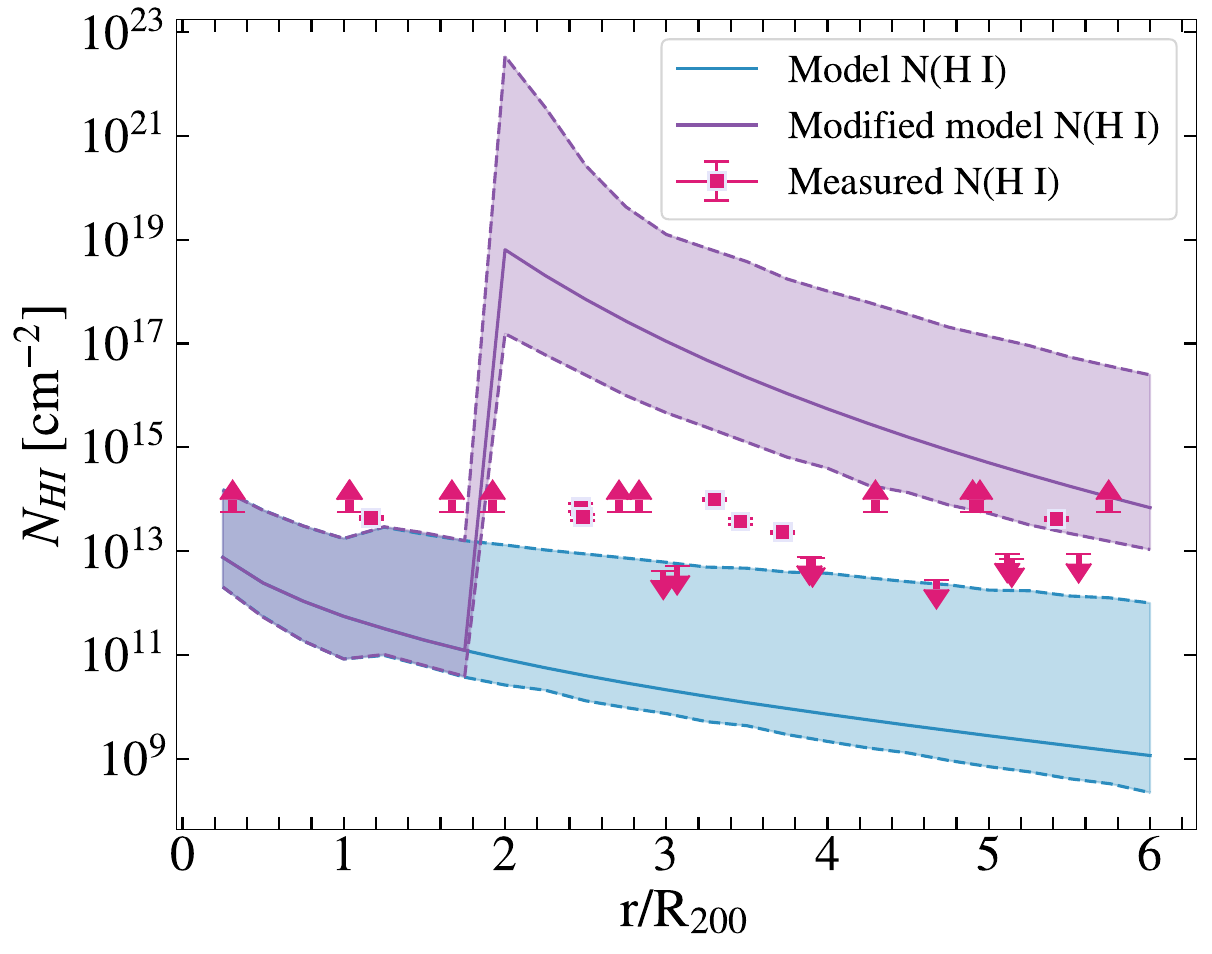} }}%
    \caption{The calculated \hone~column density as a function of \rtwo~for the toy model. The associated 1$\sigma$ error is smaller than the marker size for each measured value, shown by the solid squares. The upper and lower limits are shown by the downward- and upward-facing arrows respectively. We adopt 3$\sigma$ upper limits. The solid lines show the median of the model profiles, and the shaded regions between the dashed lines show the associated 16th and 84th percentiles. Using the \cite{lau_2015} temperature profile (solid blue line) produces a smooth, gradually decreasing profile consistent with the decrease in density we see in Figure \ref{fig:tempdens}. The modified temperature profile produces a column density enhancement at 2 \rtwo, the location of the introduced discontinuity, which drops off more quickly than the unmodified profile at higher radii (dashed purple line).  }%
    \label{fig:modelcoldens}%
\end{figure}

\section{Discussion} \label{sec:discussion}
\subsection{A higher \hone\ covering fraction}
We find that \hone\ has a higher covering fraction than \osix\ within 4 \rtwo. A potential source of this higher covering fraction for the neutral hydrogen could be due to some fraction of the \hone~in this region of the ICM having very low metallicity and therefore being fairly pristine. \cite{tripp_2005} found evidence of this in the outskirts of the Virgo Cluster, where they detect a sub-damped \hone~Lyman$\alpha$ absorber of extremely low-metallicity. It is possible that the \hone~we detect at smaller radii is less chemically enriched than the \hone~found at larger radii, and the higher covering fraction we measure for \hone~than \osix~is a reflection of that. The ability to achieve higher S/N data and detect weaker absorption components could help confirm this speculation. This may be possible with future UV observatories, such as the Habitable Worlds Observatory.

\subsection{Tentative excesses in \dndz}
We find intriguing elevation of \dndz~for \hone~at higher W$_{lim}$ and a $>$2$\sigma$ elevation of \osix~at lower W$_{lim}$ in different regions of the cluster environment. In the following, we discuss potential physical interpretations of these results, focusing on the accretion shock. We establish a phenomenological model to better interpret our findings and compare the estimated column densities with those derived from the model.

\subsection{A phenomenological model}
We derive a semi-empirical model to compare our measured column density values to self-similar models by examining N(\hone) as a function of the clustocentric radius. The following subsections detail the methodology and findings of our model. Briefly, we calculate a radially-dependent density profile from pressure and temperature profiles adopted from the literature. We calculate the density at each point through the sphere, producing a radial profile of \nhi, the \hone~volumetric number density, which we expect to drop at higher radii and in lower-density environments. Likewise, we expect the column density of \hone~to be higher in cool and dense regions where the neutral gas may survive. Next, if we consider the radial dependence of \xhi, the ionization fraction of \hone, we expect it to be lower at small radii and increase until a plateau is reached that is consistent with the \xhi~ of the IGM. We calculate the column density of \hone~at multiple impact parameters by multiplying \xhi~ and $n_H$ to get the n$_{HI}$ and integrating over a line of sight. The resultant N(\hone) profile is compared with our measured values as a function of clustocentric distance.

\subsubsection{Pressure, temperature, and density profiles}\label{subsubsec:modelprofiles}
We calculate the density profile by adapting the \cite{arnaud_2010} pressure profile and the temperature profile from \cite{lau_2015}, where we extrapolated both profiles to 6 \rtwo~-- the impact parameter limit of our observational study. Figure \ref{fig:tempdens} shows these profiles. As the mass-weighted temperature profile of \cite{lau_2015} asymptotically approaches $\sim 10^6$ K, the temperature of the warm-hot intergalactic medium (WHIM), it does not include gas at lower temperatures in the IGM. Namely, the diffuse IGM in photoionized equilibrium is expected to be commensurate with a temperature of $\sim 10^4$ K (\citealt{dave_1999,dave_2010}). For this reason, we include a modified temperature profile that drops to a constant 10$^4$ K at 2 \rtwo, and for all values beyond it, through an introduced discontinuity in the profile. The \cite{lau_2015} profile and the modified temperature profile can be seen in the top panel of Figure \ref{fig:tempdens}.

We obtain the volumetric number density of \hone, n$_{\hone}$, as follows:
\begin{equation}
    n_{HI}(r) = \chi_{HI}(r)  n_H(r)
\end{equation}
where $\chi_{\hone}$ is the neutral hydrogen fraction and $n_H$ is the volumetric number density of hydrogen. The total hydrogen number density, $n_H$, was calculated assuming the ideal gas law, where a hydrogen gas mass fraction of 0.75 was adopted for consistency with cosmological abundances (\citealt{boesgaard_1985}). The
\cite{oppenheimer_2013} non-equilibrium ionization models were used to calculate $\chi_{\hone}$ for the temperature and density values of the profiles described above. The $n_{\hone}$ profile is then used in our column density calculations, described in the following section.

\subsubsection{\hone~column density comparison}
We calculated the column density of \hone, N(\hone), along a mock quasar line of sight observation through the cluster by integrating the number density of \hone~across a 500 kpc range (Figure \ref{fig:modelcoldens}). We chose this line-of-sight distance over which to integrate N(\hone) because we want to integrate over a distance large enough to capture large column densities if they exist, while still maintaining a physical scale on par with that of the CGM of a L$_*$ galaxy, which is thought to be the origin of most strong Ly-$\alpha$ and \osix~absorbers (\citealt{prochaska_2011}). N(\hone) was calculated for varying impact parameters spanning $0.25 \leq$ r/\rtwo~$\leq 6$. The model using the unmodified temperature profile follows a steady decline in \hone~column density as the distance from the cluster increases. However, the modified temperature model produces a jump in the \hone~column density at the discontinuity of the introduced temperature of $10^4$ K. 

The pink markers in Figure \ref{fig:modelcoldens} show our estimated N(\hone) values, with squares depicting measured values, and upper and lower limits represented by downward- and upward-facing arrows respectively. The lower limits and point within 1 \rtwo~reveal estimated column densities that are greater than both models. Beyond \rtwo~we see that the N(\hone) values are not consistent with either model but are bracketed by the column density profiles derived from both the cool (10$^4$ K) and warm-hot (10$^{5-6}$) gas. Lower limits lie above the warm-hot gas model, while upper limits lie below the cool gas model. Here, we have modeled two extreme cases of either wholly cool or wholly warm-hot gas constituting the ICM in cluster outskirts. However, our measured N(\hone) values may suggest a multiphase mixture of gas in this region since the measured column density values appear to lie outside the predictions and their respective 16th and 84th percentiles for these two models (see Figure \ref{fig:modelcoldens}). As the two models bracket the upper and lower limits, the likelihood of a multiphase solution is further supported as opposed to an intermediate temperature solution since multiphase gas is necessary to reconcile values across the density range covered. However, our measurements are consistent with the models when we compare them to 97.5\% of the modeled data for each profile, as opposed to 68\% represented by the 16th and 84th percentiles. The percentiles corresponding to 97.5\% of the data (2.5 and 97.5) are not included in the figure for clarity, as the values increase substantially when we include uncertainties for $>$68\% of the modeled data. However, as shown above, our results find the \dndz~values in the cluster outskirts are consistent with the field value (Figure \ref{fig:HI_dndz}). Thus, our data reveal some complexity not captured by the simplified models, particularly in the ICM-IGM interface regions, and it is important to consider our results in the context of a range of physical conditions to achieve better constraints for current and future modeling.

\subsection{Accretion shock: source of elevated \hone?}
As discussed in Section \ref{sec:intro}, clusters are known to contain `internal' virial shocks and an `external' accretion shock (\citealt{molnar_2009}). This outer accretion shock is identified as a discontinuity in the entropy profile of the cluster (\citealt{lau_2015,shi_2016,zinger_2018b,aung_2021}), acting as a boundary of phase change in the ICM. Therefore, a neutral gas will undergo shock-heated ionization upon interaction with the shock front. The location of the slight \hone~elevation in our data coincides with the location of the accretion shock identified in simulations (\citealt{molnar_2009,zinger_2018a}). It also coincides with a pressure profile deficit attributed to shocks in recent SZ observations (\citealt{anbajagane_2022}). Our study covers impact parameters from within to well beyond the accretion shock and thus may identify observational indicators of the shock's presence. Therefore, it is possible that the slightly elevated \dndz~of \hone~we detect between 2--3 \rtwo~is consistent with a scenario in which a buildup of \hone~occurs at the cluster's outer accretion shock, producing a higher total number density of hydrogen (see bottom panel of Figure \ref{fig:tempdens}). 
 
We interpret this suggestive elevation as arising from infalling gaseous media (CGM and/or IGM) that builds up before crossing the shock boundary and experiencing ionization, creating a shell of neutral hydrogen. A schematic depicting this physical scenario is shown in Figure \ref{fig:diagram}, which shows a slice of an idealized, spherically symmetric cluster. The \hone~buildup is depicted by the purple `cloudlets' clustered around the accretion shock, with the IGM, the cluster core, and the virial shock labeled for reference. The quasar sightlines pass through the cluster at different radii, with the relative \hone~incidence of absorbers expected from that sightline depicted in the plot below the cluster. Within \rtwo, the gas temperature is $\sim10^{6.8} - 10^8$ K, which means that neutral hydrogen must have a very low ion fraction to exist. As we move outside \rtwo~and the temperature drops, the neutral fraction increases, and \hone~can more readily survive. This \hone~outside the virialized extent of the cluster can enhance clumping that simulations unanimously predict at the outskirts (\citealt{nagai_2011}, \citealt{walker_2019}). These clumps are density enhancements that contribute to the multiphase conditions of the ICM in the outskirts and can be traced via absorption line spectroscopy. 

In the outskirts, beyond 2--3 \rtwo, we expect that the number of \hone~absorbers detected should be consistent with the field value. If a buildup occurs near the shock front, it can be expected that sightlines piercing this region will contain more \hone~due to the higher chance of absorption in a sightline that passes through the limb of a shock front. This is not to be confused with the lower incidence of \hone~\textit{within} the shock front following ionization, as opposed to the higher incidence along the limb of the shock front itself. The higher probability of detecting absorption along the limb is due to the dense grouping of \hone~clouds along the outside of the front. Thus, we expect a higher chance of \hone~detection at the shock front, where the amount of absorption present is elevated above the ambient field value of \hone~absorption associated with the IGM. Within the shock front, where the ICM temperature continues to rise as one gets closer to the cluster's core, we expect neutral hydrogen to be ionized and to detect fewer \hone~absorbers. The bottom panel of Figure \ref{fig:diagram} shows the observational signature that we expect to be consistent with this scenario, where the amount of \hone~absorbers detected beyond the shock front is consistent with the field value, the amount of \hone~rises as we pierce the shock front, and decreases within this region as sightlines probe closer to the core. This signature is reflected in our \dndz~trend, and we thus interpret this result as the statistical detection of the outermost accretion shock front.

\subsection{Comparison to Previous Studies}
Similar recent work from \cite{mishra_2024} analyzing \hone~Ly$\alpha$, \osix, and \cfour~in a single stacked spectrum of clusters probed by QSOs finds no \osix~detections around their cluster sample sourced from seven cluster catalogs. This may appear in tension with our results, however, our study adopts selection criteria that place our sample in a more constrained redshift and mass range. Our sample, firmly in the cluster mass regime, has M$_{500c}$ values which range from 0.87--10 $\times 10^{14}$ M$_\odot$ (equivalent to M$_{200c} \geq 10^{14}$ M$_\odot$), while the sample in \cite{mishra_2024} has M$_{500c}$ values that range from 0.2--12.9 $\times 10^{14}$ M$_\odot$ (equivalent to M$_{200c} \gtrsim 0.28 \times 10^{14}$ M$_\odot$), thus, it includes galaxy groups. Additionally, their redshift range spans 0.01--1.1 with a median redshift of 0.44, while our sample has a redshift range of $\sim$0.1--0.3, with a median redshift of 0.16, chosen to mitigate redshift evolution effects in the cluster environment. 

Our cluster identification is consistently drawn from a single source catalog instead of the seven catalogs from their sample. We adopted a less rigid velocity window $\pm1500$~km s$^{-1}$ for identified absorbers in contrast to the variable window $\pm500$ and $\pm300$~km s$^{-1}$ adopted in their work for Ly-$\alpha$ and \cfour~respectively. In addition, we identify all absorbers in each quasar spectrum in our sample and do not adopt a stacked spectral analysis. Our spectrum-by-spectrum analysis allows us to confidently identify our \osix~absorbers, of which there are admittedly few, and all of which contain an accompanying \hone~detection. We suspect that the stacked spectrum used for analysis in \cite{mishra_2024} may contain significant enough contamination to prevent a statistically significant detection of \osix. 

One source of this contamination would arise from the inclusion of group-mass halos, whose halo sizes and thermodynamic states are significantly different enough from those of cluster halos that \osix~is expected at a much lower impact parameter than the median $\rho$ = 5 Mpc of their sample. Another potential source of contamination that could prevent a detection of \osix~in their study is the broad $z$ their sample spans, which introduces potential redshift evolution effects in the cluster environment that could likewise impact where \osix~is expected to be detected. The inclusion of these regions into the stacked spectrum used for analysis can be preventing a statistically significant \osix~detection from appearing. The scope of our analysis is thus distinct from that of \cite{mishra_2024}, and our findings can be viewed in the context of their results rather than opposed to or in contrast with them. Future studies of \osix~in the cluster environment, namely to high clustocentric radii, will bolster the low statistics we currently have of detected \osix~absorbers in this regime.

\cite{muzahid_2017} detect what are predicted to be a sub-damped Ly$\alpha$ absorber (sub-DLA; log N(\hone)/cm$^{−2}$ $\sim$ 19.3) and partial Lyman limit systems (pLLS; N(\hone) $>$ 10$^{16.2}$ cm$^{−2}$) around three clusters. Two of the detected \hone~absorbers are located at clustocentric impact parameters between 2.8--3.5 \rtwo, which are consistent with the slight elevation detected between 2--3 \rtwo~in our study. As the slight elevation we detect is not statistically significant, it is imperative that future studies target cluster-QSO systems probed in this region for confirmation of the excess in \hone, if there is one. 

Recent work from \cite{anand_2022} uses \mgtwo~absorption as a tracer of cold gas contents in galaxy cluster cores (within R$_{500}$) and observes the strength of \mgtwo~absorption decreasing as the clustocentric distance increases. They find that the covering fraction of weak absorbers is higher than that of strong absorbers at the same distance, is higher in clusters that contain a larger fraction of star-forming galaxies, and is higher in sightlines located closer to satellites than not. This alludes to a connection between the presence of cool gas and satellites in the cluster environment, especially star-forming ones. \citealt{anand_2022} shows that the gas traced by \mgtwo~absorption in their study is dominated by cold cloud complexes within the ICM, and this gas is likely circumgalactic in origin, initially entering the ICM in the CGM of infalling satellites that experience stripping. This scenario is favored over the gas arising from the stripped ISM of infalling satellites. In the context of our results, the cool \mgtwo~detected within R$_{500}$ likely arising from the cold CGM removed from the satellites aligns with our picture of an accretion shock as a mechanism of cold gas removal for infalling cluster members. As \mgtwo~has been shown to have a higher detection rate within the virial radius of clusters (\citealt{lee_2021a}), and we detect \hone~from 1 to 6 \rtwo, we expect that, at least within the virial radius, there should be sufficient \mgtwo~for detection in our cluster sample. We expect that the strength of absorption will decrease with increasing clustocentric distance however, so in the outskirts ($>$ 2\rtwo), there may not be a high enough incidence of \mgtwo~for detection.

\cite{tejos_2016}, from which we adapt our methodology for absorption line identification, examines the \dndz~of \hone~and \osix~with regard to pairs of filaments connecting cluster pairs. While their work uses the same metric as ours (\dndz), our study is set up such that analysis is performed in relation to a clustocentric impact parameter, and the values they find are thus not directly comparable to our study. They find a tentative excess of total \hone~--- $\sim$2 times the field value and detected to a 1$\sigma$ confidence level --- that starts at 3 Mpc relative to the filament connecting the cluster pair. They attribute this excess primarily to broad Ly$\alpha$ absorbers (BLAs; Doppler parameter b $\geq$ 50 km s$^{-1}$). In contrast to the narrow \hone~absorbers that likely trace cool, photoionized gas, BLAs are found in warm-hot (T $\geq 10^5$ K) gas. As the gas temperature rises, thermally-broadened line profiles are produced through collisions in the diffuse neutral hydrogen in this temperature regime (\citealt{tepper-garcia_2012}). This typically makes the absorption signatures shallow and weak, making them harder to detect in lower S/N quasar spectra. This thermally broadened \hone~has been predicted to arise from shock-heating in WHIM filaments (\citealt{tepper-garcia_2012,richter_2006}). 

\osix~is another ion expected to originate from shock-heating occurring in the WHIM, as discussed in Section \ref{subsec:osixinterp}. \cite{tejos_2016} also detect a tentative \osix~excess of $\sim$4 times the field expectation, to a 1$\sigma$ confidence level, at 3 Mpc relative to the filaments connecting the cluster pairs. Their detected excess peaks at 3 Mpc and remains at or above the field value until $\sim$5 Mpc, beyond which the values agree with the field. While the origins of these detected \hone~and \osix~excesses may appear to arise from the same mechanism due to their overlap in location, \cite{richter_2006} purports that it is questionable whether the shock-heated \hone~and \osix~in the WHIM are tracing the same gas. Therefore, while shock-heating is plausibly the mechanism giving rise to these detections, the shocks that cause the excess of \hone~and \osix~are most likely different in origin (accretion shock- vs. WHIM-originated respectively).

\subsection{\osix~in the outskirts of clusters}
\label{subsec:osixinterp}
Within the putative accretion shock region, we measure a \dndz~for \osix~that is statistically consistent with the field value (within 1$\sigma$). Within \rtwo, we might expect some \osix~absorption due to the multiphase nature of the ICM, e.g., associated with cooling flows. In the CGM of star-forming galaxies, \cite{stern_2019} find consistency between predicted and observed \osix~column densities for cooling flow models in which some mechanism, such as feedback, preferentially heats the gas in the outer halo to temperatures conducive to \osix~production. Another such origin for \osix~in this region is a cooling flow that develops when the gas beyond the innermost accretion shock free-falls into the core (\citealt{stern_2018}), not unlike the accretion of infalling gas and satellites to the cluster core, which is the source of the internal virial shock in clusters. Although our study is focused on the outskirts of clusters, which is a different environment than the context of the cooling flow models, cooling flows are known to exist in cluster environments, as evidenced by the strong correlation of filamentary H$\alpha$ flux and X-ray cooling flow rate and the physical coincidence of H$\alpha$ emitting filaments and the cool X-ray gas substructure in clusters (\citealt{cowie_1980,fabian_2003,crawford_2005,mcdonald_2010,jimenez-gallardo_2022}). If \osix~is produced via cooling flows in the cores of these clusters and makes its way to the `outer' halo via outflow/feedback processes, it is plausible as a potential mechanism producing the tentative elevation of \osix~we detect from 1--2 \rtwo.

Our observations indicate a dearth of \osix~between 2--4 \rtwo~(see Figure \ref{fig:OVI_EW}), which is partially coincident with the tentative excess in \hone~we detect from 2--3 \rtwo. In the context of the outer accretion shock, it is possible that the high entropy gas ionizes any oxygen beyond \osix~through shock heating, and we therefore see an apparent lack of \osix~in this region. 

As we approach the far outskirts of the clusters, we detect a $>$ 2$\sigma$ elevation in \osix~for small limiting equivalent widths. This higher detection of \osix~in the far outskirts is highly intriguing as \osix~is often associated with the CGM in the low-z universe. In the discourse over whether \osix~absorption arises in the CGM or IGM, prior work has purported that the majority of \osix~detected via absorption is CGM-related and rarely arises from the IGM (\citealt{prochaska_2011}). To interpret our \osix~detections, we must consider whether the gas we are detecting is circumgalactic or intergalactic in origin because if the latter, this result may be a signature of the warm-hot intergalactic medium (WHIM). The primary heating mechanism for the WHIM ($10^5-10^7$ K) is thought to be gravitationally driven shocks (\citealt{dave_2001}) that arise from perturbations, which interact with the low-density filamentary structure of the cosmic web. \osix~is a favored tracer for the WHIM (\citealt{cen_2001,dave_2001,tepper-garcia_2011,tejos_2016,ahoranta_2021}), yet disentangling the ionization mechanism giving rise to \osix~absorbers can be challenging, as \osix~can be produced via collisional ionization, photoionization, or a mix of the two. The latter is especially true in the case of complex, multiphase absorbers, as shown by \cite{tripp_2008}, where over half of their non-QSO associated \osix~absorbers are multiphase systems for which both collisional ionization and photoionization are plausible ionizing mechanisms. 

If we consider the detected \osix~as circumgalactic in origin, and therefore tracing the hot CGM, it is pertinent to consider whether this is probing the outer, hot halo of the ICM, or the CGM of to-be accreted satellites at large clustocentric distances. When we factor in the galaxy density comparison, we see no clear systematic excess of galaxies in sightlines where we detect \osix~absorption, so it is unlikely this \osix~is tracing the CGM of satellites at large clustocentric distances. However, we do note the caveat that the uncertainties are large in photo-z. We do not have a definitive spectroscopic census of galaxies around these sightlines to high completeness.

The cluster environment is conducive to multiphase gas in the outskirts, as clusters, existing at the nodes of the cosmic web, are fed by cool, low-density filamentary streams from the IGM that interface the hot, dense ICM. Non-thermal pressure support and gas motion, as well as density perturbations from accreting galaxies and filaments, give rise to clumping and substructure in cluster outskirts (\citealt{nagai_2011,roncarelli_2013,vazza_2013,zhuravleva_2013,battaglia_2015,avestruz_2016a}). Out at 4--6 \rtwo, where we are comfortably beyond the cluster's outskirts and in the IGM, multiphase gas is not expected to subsist at the same density it would within the overdense cluster region. The intriguing 2 to 3$\sigma$ \osix~elevation detected in this region may be indicative of shock-heating in the WHIM. The large clustocentric distance of the detection further supports the gas being WHIM in origin.

\section{Conclusion}
\label{sec:conclusion}
In this study, we use archival HST/COS spectra of quasars to study the ICM in 26 foreground clusters (M$_{200} > 10^{14}$ \msun~and $z \sim$ 0.1--0.3), tracing \hone~and \osix~from within \rtwo~to 6 \rtwo. We identify all absorbers in each sightline before determining which absorption system(s) are associated with the cluster (within $\pm$1500 km s$^{-1}$ of the cluster systemic velocity). We then calculate W$_r$, \dndz, and the covering fraction for \hone~and \osix~as a function of the impact parameter. Our main results are as follows:
\begin{enumerate}
    \item We identify an intriguing elevation in \dndz~of \hone~from 2--3 \rtwo, detected to a 2$\sigma$ confidence level at W$_{lim}$ = 50 \& 100 m\AA. This is coincident with the clustocentric distance of recently identified strong \hone~absorbers and pressure deficits attributed to shocks in SZ profiles of clusters. Coupled with the accretion shock front acting as a boundary of phase change in the ICM, this elevation may be suggestive of a buildup of \hone~at the outer accretion shock.
    \item We establish a phenomenological model of such an \hone~buildup and find that neither self-similar models that predict warm-hot (10$^6$ K) gas in the outskirts nor cool (10$^4$ K) gas models are capable of reproducing column densities consistent with those estimated in our study. Rather, the estimated column density values are bracketed by these models, with upper limits aligning with the warm-hot model and lower limits below the cool model, implying the gas detected in the outskirts is multiphase (T $\sim$ 10$^{4-6.5}$ K) as it spans this range, and these simple thermodynamic models are not reconcilable with the true physical conditions of all clusters.
    \item We identify an excess of \osix~from 1--2 and 4--5 \rtwo, detected to a 2$\sigma$ confidence level for W$_{lim}$ = 20 \& 50 m\AA. While the primary mechanism giving rise to our \osix~detections cannot be determined conclusively, the large impact parameters at which we detect the slight elevation may indicate that the detected gas resides in the WHIM. As we see no clear systematic excess of galaxies in sightlines where we detect \osix~absorption, we deem it unlikely this gas is tracing the hot CGM (see Section \ref{subsec:osixinterp}).
    \item $\sim$25\% of our sample show metal-rich absorption systems. These systems span the full impact parameter range of our study and show no correlation with clustocentric impact parameter and galaxy density.
    \item We find no clear relation between W$_r$ and impact parameter for \osix. Likewise, the relationship between impact parameters and statistically significant detections and upper limits is scattered across the impact parameter range. 
    \item We find \hone~has a higher covering fraction than \osix~within 4 \rtwo, and the two have consistent covering fractions beyond 4 \rtwo~within their respective binomial confidence intervals. As the gas density decreases in the cluster outskirts and then further decreases in the transition to the IGM, it is not unexpected that we do not find absorption in every sightline at these large radii given the small dz in each QSO sightline.
\end{enumerate}

As mentioned above, a small number, $\sim25\%$, of systems in this study exhibit several metal line detections. We are conducting follow-up observations of these metal-rich systems by obtaining spectra for satellites within 1000 kpc of the quasar (projected). The spectra will be analyzed to investigate the possible origin(s) of the identified metals. Additionally, we plan to infer the density field of the cosmic web around the clusters in our sample, elucidating the filaments, voids, and nodes that comprise it. Our preferred method for this includes the Monte Carlo Physarum Machine (MCPM), a cosmic web reconstruction algorithm based on slime mold feeding patterns (\citealt{elek_2022}). An analysis combining the reconstruction of the cosmic web with absorption spectroscopy has already been successfully demonstrated with MCPM (\citealt{burchett_2020}), and MCPM has been shown to identify filaments to a higher fidelity than comparable cosmic web reconstruction algorithms (\citealt{hasan_2024}). We will use SDSS data in conjunction with MCPM and other available ancillary data to determine how the cosmic geometry of the cluster environment might affect identified absorption in that system, which is otherwise impossible to infer through quasar absorption line spectroscopy alone. This will also allow us to further refine the idealized assumption of spherical symmetry in the cluster environment, namely, when interpreting the nature of the accretion shock front, which is likely asymmetric. 

Future work further characterizing the physical conditions of the ICM in cluster outskirts will be crucial to refining our understanding of the ICM/CGM/IGM interface(s) and the role of the accretion shock in these environs. Quasar absorption line spectroscopy remains a powerful technique for probing the diffuse gas in this regime and providing key constraints on gas kinematics and density. Due to the low confidence level at which we detect these slight excesses in \hone~and \osix, further absorption line studies probing cluster-QSO systems at these clustocentric impact parameters are imperative for bolstering statistics and confirming the statistical significance of any potential excesses. As X-ray and SZ observations continue to progress, the capability of a multi-wavelength approach to observing the outskirts of clusters will become more readily available -- an approach that will be crucial for constraining the physical conditions of the elusive nature of the ICM in this region.

\begin{acknowledgments}

We would like to thank F. Hasan, R. Bordoloi, S. Mishra, S. Muzahid, and J. Stern for stimulating interesting and insightful conversation on this work. 
\end{acknowledgments}

\newpage
\appendix
\section{Table of all identified absorbers}
This section of the appendix includes all \hone\ $\lambda1216$ and \osix\ $\lambda1032$ absorbers within $\pm$1500 km s$^{-1}$ of the cluster's systemic velocity. Non-detections are denoted by `ND' in the rank (8th) column, where the listed W$_\lambda$ is the 3$\sigma$ upper limit. Lower limits include the measured W$_\lambda$ and reliability rank as described in Section \ref{subsec:lineid}, with the adopted column density saturation threshold, as described in Section \ref{subsec:linemeas}, listed for log (N/cm$^{-2}$).

\onecolumngrid
\begin{longtable}{cccccccc}
\caption{(1) Cluster ID from redMaPPer catalog. (2) Quasar name. (3) Ion and rest frame wavelength of the strongest transition observed for this ion. (4) Redshift of the identified absorber. (5) Velocity of the absorption component relative to systemic velocity. (6) Observed equivalent width, from apparent optical depth (see Section \ref{subsec:linemeas}), of the listed ion's strongest transition. (7) Estimated column density for the measured equivalent width (see Section \ref{subsec:linemeas}). (8) Component reliability rank (`a' reliable, `b' possible, `c' uncertain, and `ND' for non-detection; see Section \ref{subsec:lineid} \& \ref{subsec:linemeas}).  } \\
\hline
Cluster ID & QSO Name & Line & z$_{abs}$ & v$_{abs}$ ($\mathrm{km~s^{-1}}$) & W$_\lambda$ (\AA) & log (N/cm$^{-2}$) & rank \\
(1) & (2) & (3) & (4) & (5) & (6) & (7) & (8) \\
 \hline
 \hline \endfirsthead
 \hline
Cluster ID & QSO Name & Line & z$_{abs}$ & v$_{abs}$ ($\mathrm{km~s^{-1}}$) & W$_\lambda$ (\AA) & log (N/cm$^{-2}$) & rank \\
(1) & (2) & (3) & (4) & (5) & (6) & (7) & (8) \\
 \hline
 \hline \endhead
30907 & PG1259+593 & HI 1216 & 0.21 & -436 & 0.11 $\pm$ 0.01 & 13.37 $\pm$ 0.02 & b \\
30907 & PG1259+593 & OVI 1032 & 0.21 & -434 & 0.02 $\pm$ 0.0 & 13.25 $\pm$ 0.07 & b \\
10214 & 4C-13.41 & HI 1216 & 0.16 & ... & $<$ 0.04 & ... & ND \\
10214 & 4C-13.41 & OVI 1032 & 0.16 & ... & $<$ 0.05 & ... & ND \\
46 & SDSS-J104741.75+151332.2 & HI 1216 & 0.21 & -339 & 0.16 $\pm$ 0.02 & $> $13.75 $\pm$ 0.3 & b \\
46 & SDSS-J104741.75+151332.2 & HI 1216 & 0.21 & -389 & 0.14 $\pm$ 0.02 & 13.73 $\pm$ 0.14 & a \\
46 & SDSS-J104741.75+151332.2 & OVI 1032 & 0.21 & 0 & 0.02 $\pm$ 0.01 & 0.0 $\pm$ 0.73 & c \\
33547 & FIRST-J020930.7-043826 & HI 1216 & 0.14 & -492 & 0.04 $\pm$ 0.01 & 12.93 $\pm$ 0.09 & b \\
33547 & FIRST-J020930.7-043826 & HI 1216 & 0.14 & -780 & 0.33 $\pm$ 0.01 & $> $13.75 $\pm$ 0.3 & b \\
33547 & FIRST-J020930.7-043826 & HI 1216 & 0.14 & -854 & 0.1 $\pm$ 0.01 & 13.48 $\pm$ 0.04 & b \\
33547 & FIRST-J020930.7-043826 & HI 1216 & 0.14 & 999 & 0.53 $\pm$ 0.01 & $> $13.75 $\pm$ 0.3 & c \\
33547 & FIRST-J020930.7-043826 & OVI 1032 & 0.14 & 0 & 0.01 $\pm$ 0.0 & 14.94 $\pm$ 0.31 & c \\
12649 & FIRST-J020930.7-043826 & HI 1216 & 0.31 & ... & $<$ 0.05 & ... & ND \\
12649 & FIRST-J020930.7-043826 & OVI 1032 & 0.31 & ... & $<$ 0.02 & ... & ND \\
2691 & FBQSJ1519+2838 & HI 1216 & 0.13 & 546 & 0.37 $\pm$ 0.0 & $> $13.75 $\pm$ 0.3 & a \\
2691 & FBQSJ1519+2838 & OVI 1032 & 0.13 & 452 & 0.06 $\pm$ 0.01 & 13.78 $\pm$ 0.05 & a \\
2691 & FBQSJ1519+2838 & HI 1216 & 0.13 & 444 & 0.35 $\pm$ 0.0 & $> $13.75 $\pm$ 0.3 & a \\
8791 & PG1222+216 & HI 1216 & 0.14 & -959 & 0.11 $\pm$ 0.0 & 13.44 $\pm$ 0.02 & a \\
8791 & PG1222+216 & HI 1216 & 0.14 & -1034 & 0.07 $\pm$ 0.01 & 13.15 $\pm$ 0.04 & a \\
8791 & PG1222+216 & OVI 1032 & 0.14 & 0 & 0.03 $\pm$ 0.01 & 0.0 $\pm$ 0.26 & c \\
77073 & PG1222+216 & OVI 1032 & 0.24 & -863 & 0.22 $\pm$ 0.0 & 14.62 $\pm$ 0.05 & c \\
77073 & PG1222+216 & HI 1216 & 0.24 & -892 & 0.35 $\pm$ 0.01 & $> $13.75 $\pm$ 0.3 & a \\
82403 & PG1222+216 & HI 1216 & 0.22 & ... & $<$ 0.02 & ... & ND \\
82403 & PG1222+216 & OVI 1032 & 0.22 & ... & $<$ 0.02 & ... & ND \\
24201 & QSOB1612+266 & HI 1216 & 0.13 & 1420 & 0.02 $\pm$ 0.0 & 12.64 $\pm$ 0.09 & b \\
24201 & QSOB1612+266 & HI 1216 & 0.13 & 1384 & 0.04 $\pm$ 0.01 & 12.95 $\pm$ 0.06 & b \\
24201 & QSOB1612+266 & HI 1216 & 0.13 & 1328 & 0.06 $\pm$ 0.01 & 13.12 $\pm$ 0.04 & b \\
24201 & QSOB1612+266 & HI 1216 & 0.13 & 1262 & 0.06 $\pm$ 0.0 & 13.13 $\pm$ 0.04 & b \\
24201 & QSOB1612+266 & HI 1216 & 0.13 & 832 & 0.1 $\pm$ 0.01 & 13.3 $\pm$ 0.04 & b \\
24201 & QSOB1612+266 & OVI 1032 & 0.13 & 500 & 0.03 $\pm$ 0.02 & 13.39 $\pm$ 0.27 & b \\
24201 & QSOB1612+266 & HI 1216 & 0.13 & 489 & 0.1 $\pm$ 0.0 & 13.35 $\pm$ 0.03 & b \\
24201 & QSOB1612+266 & HI 1216 & 0.13 & 439 & 0.05 $\pm$ 0.0 & 12.99 $\pm$ 0.05 & b \\
24201 & QSOB1612+266 & HI 1216 & 0.13 & 406 & 0.03 $\pm$ 0.01 & 12.81 $\pm$ 0.07 & b \\
843 & QSOB1612+266 & OVI 1032 & 0.19 & 647 & 0.09 $\pm$ 0.01 & 13.96 $\pm$ 0.03 & a \\
843 & QSOB1612+266 & OVI 1032 & 0.19 & 590 & 0.14 $\pm$ 0.0 & 14.32 $\pm$ 0.03 & a \\
843 & QSOB1612+266 & HI 1216 & 0.19 & 580 & 0.79 $\pm$ 0.01 & $> $13.75 $\pm$ 0.3 & a \\
843 & QSOB1612+266 & OVI 1032 & 0.19 & 554 & 0.18 $\pm$ 0.0 & 14.47 $\pm$ 0.02 & a \\
843 & QSOB1612+266 & HI 1216 & 0.19 & -107 & 0.03 $\pm$ 0.01 & 12.83 $\pm$ 0.08 & b \\
11725 & PG1121+422 & HI 1216 & 0.19 & 573 & 0.6 $\pm$ 0.02 & $> $13.75 $\pm$ 0.3 & a \\
11725 & PG1121+422 & OVI 1032 & 0.19 & 557 & 0.02 $\pm$ 0.0 & 13.27 $\pm$ 0.1 & c \\
11725 & PG1121+422 & HI 1216 & 0.2 & -687 & 0.06 $\pm$ 0.02 & 13.17 $\pm$ 0.16 & b \\
4241 & SBS0956+509 & HI 1216 & 0.13 & -353 & 0.34 $\pm$ 0.01 & $> $13.75 $\pm$ 0.3 & a \\
4241 & SBS0956+509 & OVI 1032 & 0.13 & 0 & 0.02 $\pm$ 0.01 & 0.0 $\pm$ 0.79 & c \\
163 & 2MASS-J14312586+2442203 & HI 1216 & 0.13 & 1042 & 0.06 $\pm$ 0.01 & 13.08 $\pm$ 0.05 & b \\
163 & 2MASS-J14312586+2442203 & HI 1216 & 0.13 & 998 & 0.08 $\pm$ 0.01 & 13.24 $\pm$ 0.04 & b \\
163 & 2MASS-J14312586+2442203 & HI 1216 & 0.13 & 964 & 0.06 $\pm$ 0.01 & 13.14 $\pm$ 0.04 & b \\
163 & 2MASS-J14312586+2442203 & OVI 1032 & 0.14 & 0 & 0.03 $\pm$ 0.01 & 0.0 $\pm$ 0.25 & c \\
11497 & 2XMM-J100420.0+051300 & HI 1216 & 0.12 & 1452 & 0.03 $\pm$ 0.01 & 12.7 $\pm$ 0.14 & a \\
11497 & 2XMM-J100420.0+051300 & HI 1216 & 0.12 & 1364 & 0.19 $\pm$ 0.01 & $> $13.75 $\pm$ 0.3 & a \\
11497 & 2XMM-J100420.0+051300 & HI 1216 & 0.12 & 118 & 0.37 $\pm$ 0.01 & $> $13.75 $\pm$ 0.3 & b \\
11497 & 2XMM-J100420.0+051300 & OVI 1032 & 0.12 & 0 & 0.05 $\pm$ 0.02 & 0.0 $\pm$ 1.11 & c \\
47627 & ICRF-J122011.8+020342 & OVI 1032 & 0.19 & 1126 & 0.04 $\pm$ 0.01 & 13.58 $\pm$ 0.08 & a \\
47627 & ICRF-J122011.8+020342 & HI 1216 & 0.19 & 1065 & 0.24 $\pm$ 0.01 & $> $13.75 $\pm$ 0.3 & b \\
47627 & ICRF-J122011.8+020342 & OVI 1032 & 0.19 & 1024 & 0.15 $\pm$ 0.01 & 14.23 $\pm$ 0.03 & a \\
47627 & ICRF-J122011.8+020342 & HI 1216 & 0.19 & 951 & 0.19 $\pm$ 0.01 & $> $13.75 $\pm$ 0.3 & b \\
47627 & ICRF-J122011.8+020342 & OVI 1032 & 0.19 & 951 & 0.05 $\pm$ 0.01 & 13.72 $\pm$ 0.05 & b \\
9713 & ICRF-J122011.8+020342 & HI 1216 & 0.23 & ... & $<$ 0.02 & ... & ND \\
9713 & ICRF-J122011.8+020342 & OVI 1032 & 0.23 & ... & $<$ 0.02 & ... & ND \\
37036 & J102512.86+480853.2 & HI 1216 & 0.14 & ... & $<$ 0.04 & ... & ND \\
37036 & J102512.86+480853.2 & OVI 1032 & 0.14 & ... & $<$ 0.04 & ... & ND \\
6097 & J102512.86+480853.2 & HI 1216 & 0.13 & ... & $<$ 0.03 & ... & ND \\
6097 & J102512.86+480853.2 & OVI 1032 & 0.13 & ... & $<$ 0.04 & ... & ND \\
337305 & SDSSJ123304.05-003134.1 & OVI 1032 & 0.32 & 815 & 0.1 $\pm$ 0.01 & 14.07 $\pm$ 0.04 & a \\
337305 & SDSSJ123304.05-003134.1 & HI 1216 & 0.32 & 803 & 0.2 $\pm$ 0.01 & $> $13.75 $\pm$ 0.3 & a \\
337305 & SDSSJ123304.05-003134.1 & HI 1216 & 0.32 & 733 & 0.25 $\pm$ 0.01 & $> $13.75 $\pm$ 0.3 & a \\
337305 & SDSSJ123304.05-003134.1 & OVI 1032 & 0.32 & 735 & 0.2 $\pm$ 0.01 & 14.43 $\pm$ 0.03 & a \\
337305 & SDSSJ123304.05-003134.1 & HI 1216 & 0.32 & 669 & 0.25 $\pm$ 0.01 & $> $13.75 $\pm$ 0.3 & a \\
337305 & SDSSJ123304.05-003134.1 & OVI 1032 & 0.32 & 651 & 0.21 $\pm$ 0.01 & 14.45 $\pm$ 0.03 & a \\
337305 & SDSSJ123304.05-003134.1 & HI 1216 & 0.32 & 589 & 0.29 $\pm$ 0.01 & $> $13.75 $\pm$ 0.3 & a \\
4411 & SDSSJ123304.05-003134.1 & OVI 1032 & 0.19 & -247 & 0.04 $\pm$ 0.01 & 13.55 $\pm$ 0.12 & b \\
4411 & SDSSJ123304.05-003134.1 & HI 1216 & 0.19 & -253 & 0.13 $\pm$ 0.02 & 13.6 $\pm$ 0.1 & b \\
4411 & SDSSJ123304.05-003134.1 & HI 1216 & 0.19 & -314 & 0.11 $\pm$ 0.02 & 13.47 $\pm$ 0.12 & b \\
9432 & J101730.98+470225.0 & HI 1216 & 0.17 & -794 & 0.11 $\pm$ 0.01 & 13.47 $\pm$ 0.06 & b \\
9432 & J101730.98+470225.0 & HI 1216 & 0.17 & -844 & 0.07 $\pm$ 0.01 & 13.18 $\pm$ 0.09 & b \\
9432 & J101730.98+470225.0 & OVI 1032 & 0.17 & 0 & 0.04 $\pm$ 0.01 & 0.0 $\pm$ 2.39 & c \\
5859 & J101730.98+470225.0 & HI 1216 & 0.15 & 661 & 0.05 $\pm$ 0.01 & 13.06 $\pm$ 0.1 & b \\
5859 & J101730.98+470225.0 & HI 1216 & 0.15 & 584 & 0.07 $\pm$ 0.01 & 13.16 $\pm$ 0.09 & b \\
5859 & J101730.98+470225.0 & HI 1216 & 0.15 & 498 & 0.05 $\pm$ 0.01 & 13.05 $\pm$ 0.11 & b \\
5859 & J101730.98+470225.0 & OVI 1032 & 0.15 & 0 & 0.05 $\pm$ 0.02 & 13.43 $\pm$ 0.22 & c \\
43551 & SDSSJ133300.83+451809.0 & HI 1216 & 0.16 & 443 & 0.24 $\pm$ 0.01 & $> $13.75 $\pm$ 0.3 & a \\
43551 & SDSSJ133300.83+451809.0 & HI 1216 & 0.16 & 338 & 0.33 $\pm$ 0.01 & $> $13.75 $\pm$ 0.3 & a \\
43551 & SDSSJ133300.83+451809.0 & HI 1216 & 0.16 & -239 & 0.2 $\pm$ 0.01 & $> $13.75 $\pm$ 0.3 & a \\
43551 & SDSSJ133300.83+451809.0 & HI 1216 & 0.16 & -424 & 0.05 $\pm$ 0.01 & 13.05 $\pm$ 0.1 & b \\
43551 & SDSSJ133300.83+451809.0 & HI 1216 & 0.16 & 1290 & 0.11 $\pm$ 0.01 & 13.41 $\pm$ 0.05 & b \\
43551 & SDSSJ133300.83+451809.0 & HI 1216 & 0.16 & 1191 & 0.13 $\pm$ 0.01 & 13.43 $\pm$ 0.05 & b \\
43551 & SDSSJ133300.83+451809.0 & OVI 1032 & 0.16 & 0 & 0.04 $\pm$ 0.01 & 13.58 $\pm$ 0.13 & c \\
7048 & QSO-B0923+201 & HI 1216 & 0.12 & ... & $<$ 0.04 & ... & ND \\
7048 & QSO-B0923+201 & OVI 1032 & 0.12 & ... & $<$ 0.04 & ... & ND \\
8183 & SDSSJ161916.54+334238.4 & HI 1216 & 0.12 & ... & $<$ 0.03 & ... & ND \\
8183 & SDSSJ161916.54+334238.4 & OVI 1032 & 0.12 & ... & $<$ 0.04 & ... & ND \\
    \hline
  \label{tab:absorbers}
\end{longtable}

\newpage
\section{Galaxy density and spectra for all systems in sample}
This section of the appendix contains the 2D galaxy density plots and spectra, like that shown in Figure \ref{fig:metalrich}, for the remaining systems in our sample. All identified absorbers for a system are included in the stacked spectra, with systemic velocity at the redshift of the cluster.
\begin{figure*}[h]
  \centering
  \includegraphics[width=\textwidth]{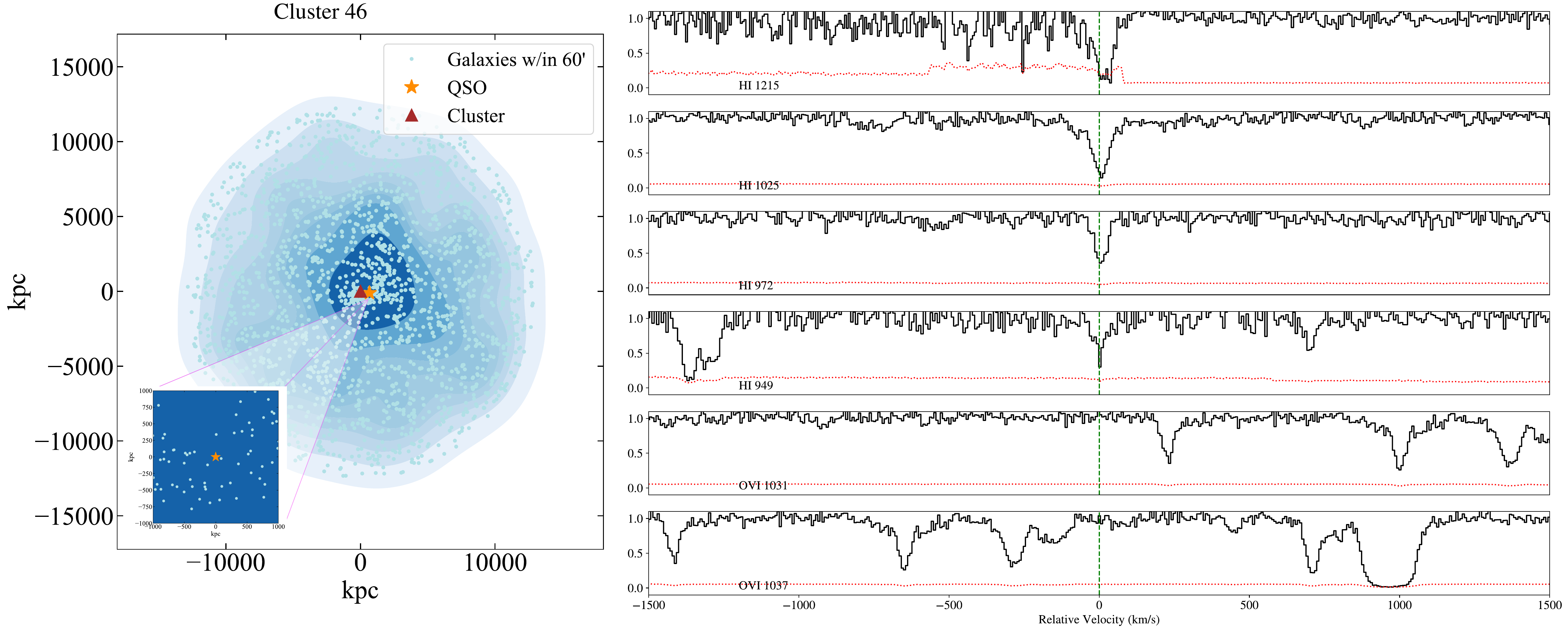}
\end{figure*}
\begin{figure*}[h]
  \centering
  \includegraphics[width=\textwidth]{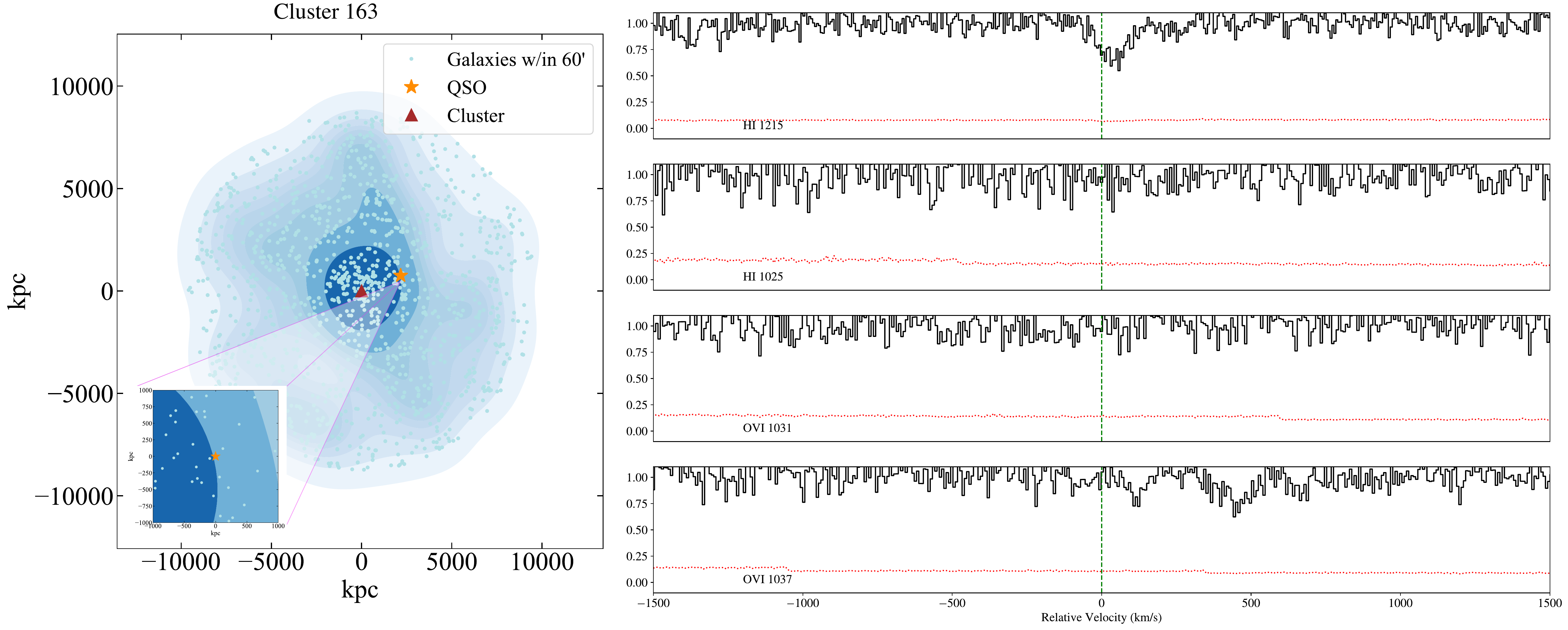}
\end{figure*}
\begin{figure*}
  \centering
  \includegraphics[width=\textwidth]{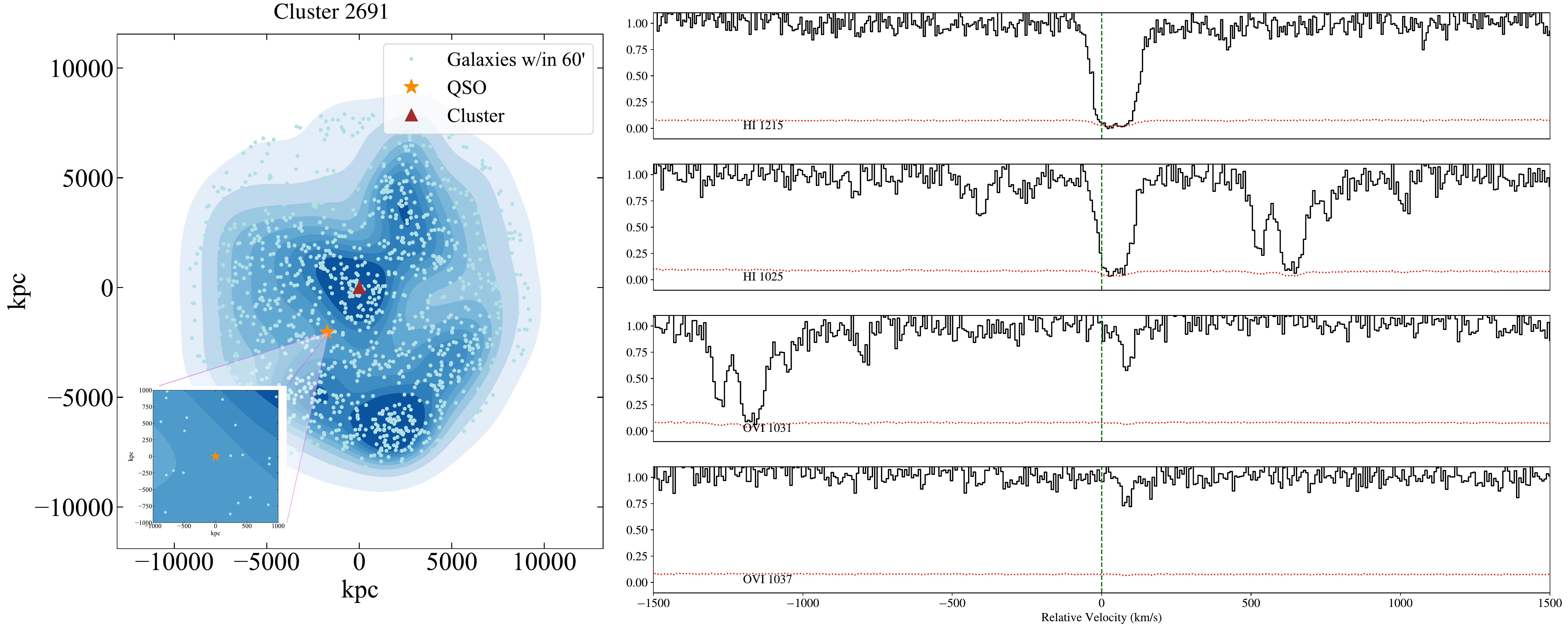}
\end{figure*}
\begin{figure*}
  \centering
  \includegraphics[width=\textwidth]{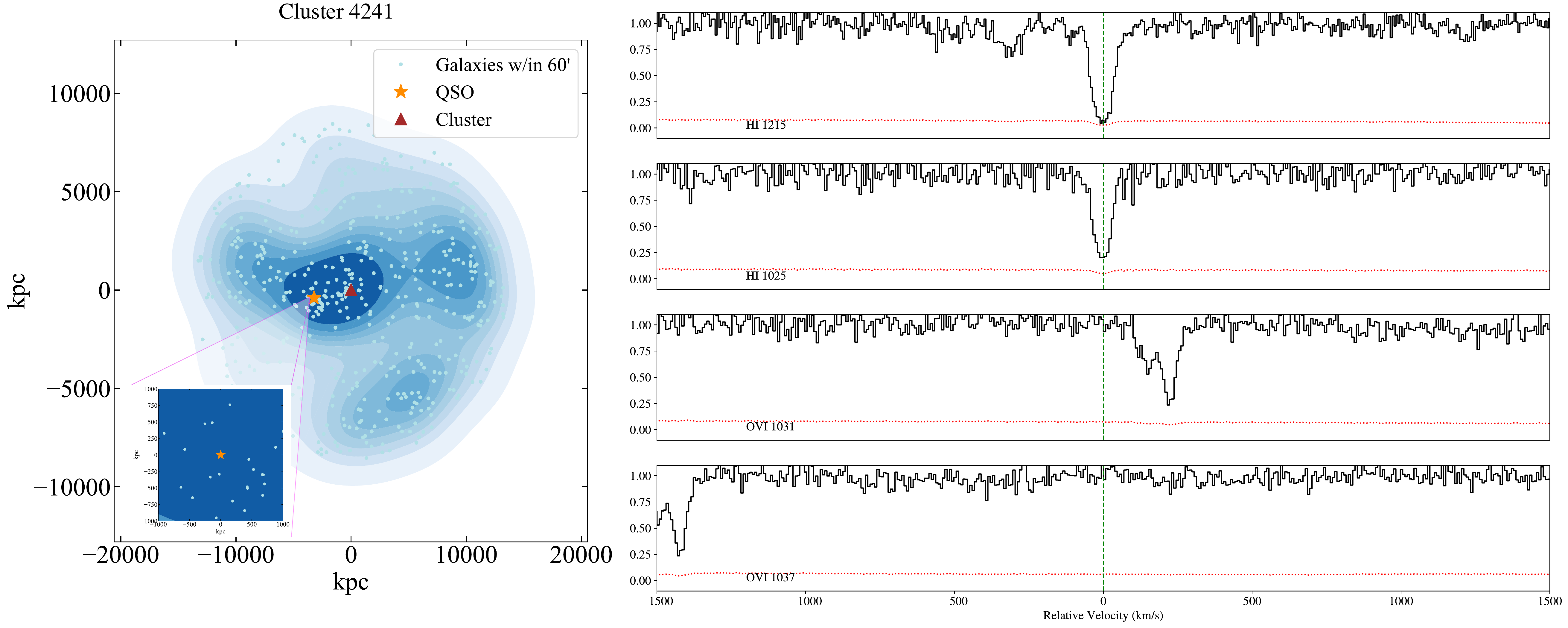}
\end{figure*}
\begin{figure*}
  \centering
  \includegraphics[width=\textwidth]{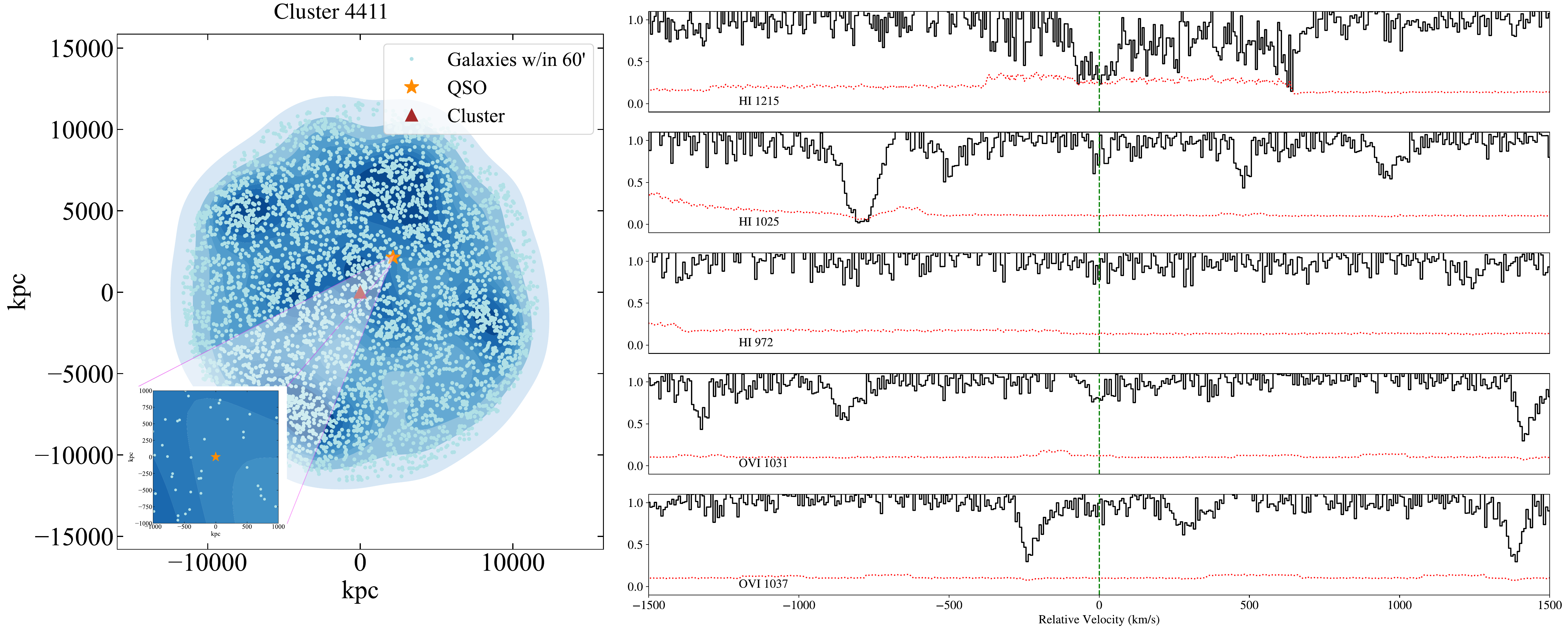}
\end{figure*}
\begin{figure*}
  \centering
  \includegraphics[width=\textwidth]{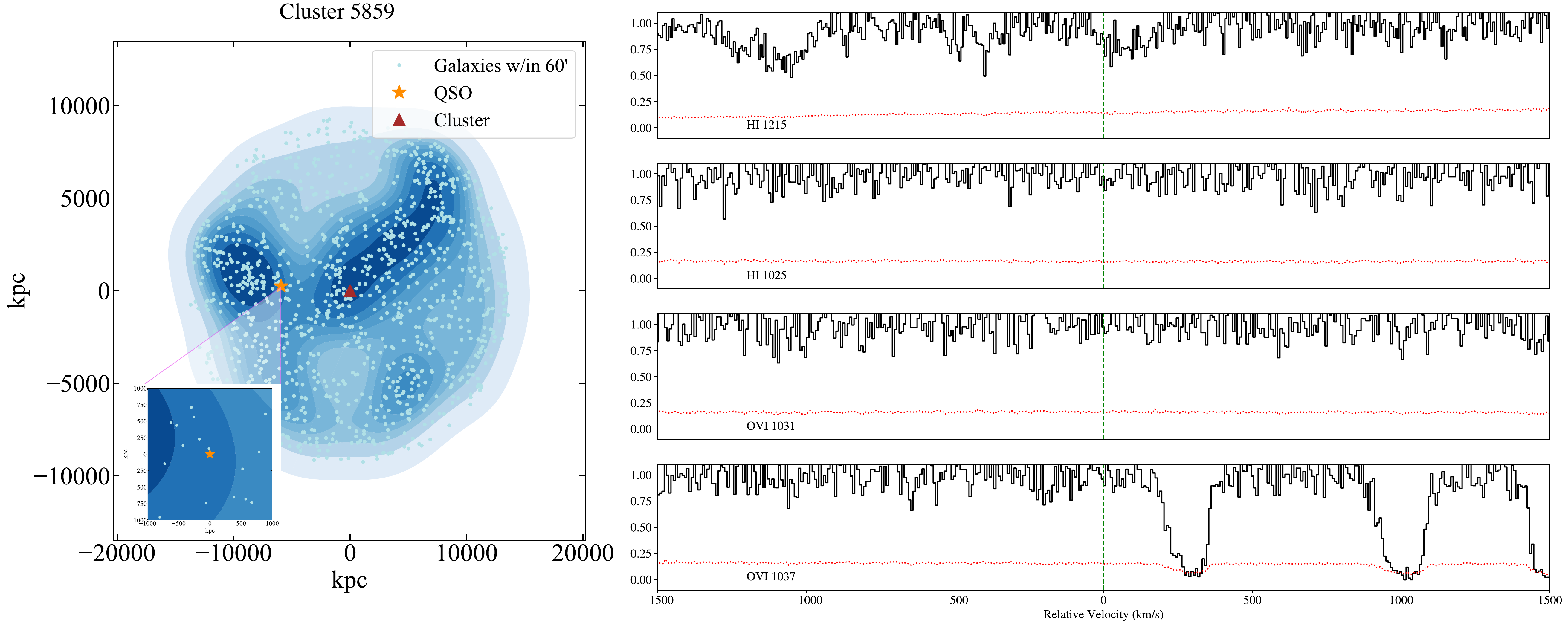}
\end{figure*}
\begin{figure*}
  \centering
  \includegraphics[width=\textwidth]{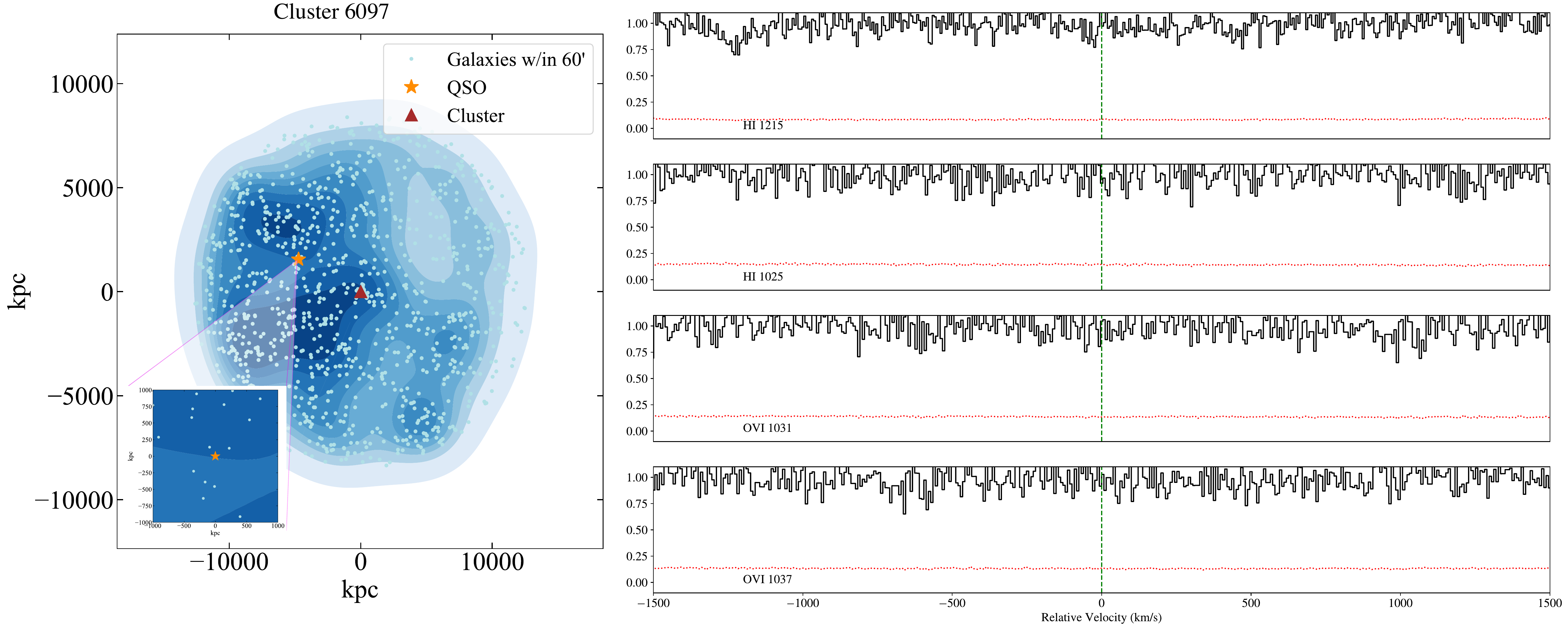}
\end{figure*}
\begin{figure*}
  \centering
  \includegraphics[width=\textwidth]{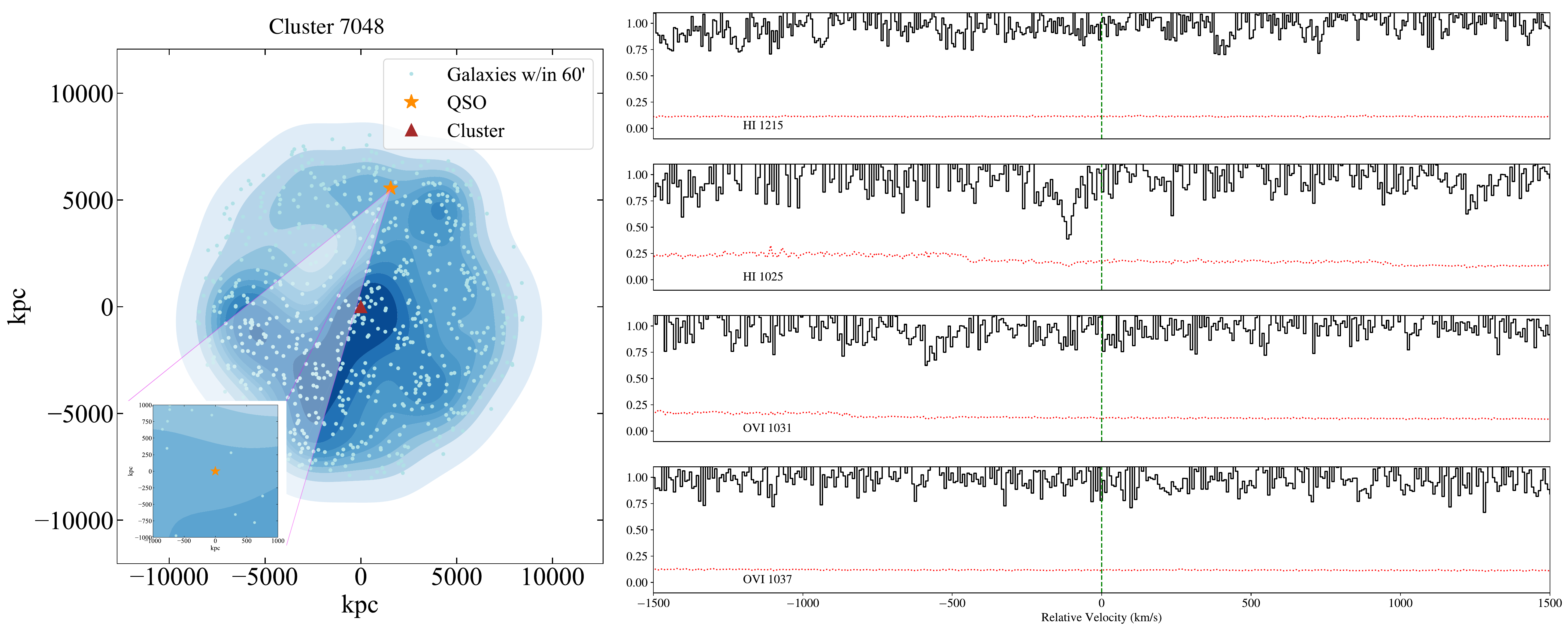}
\end{figure*}
\begin{figure*}
  \centering
  \includegraphics[width=\textwidth]{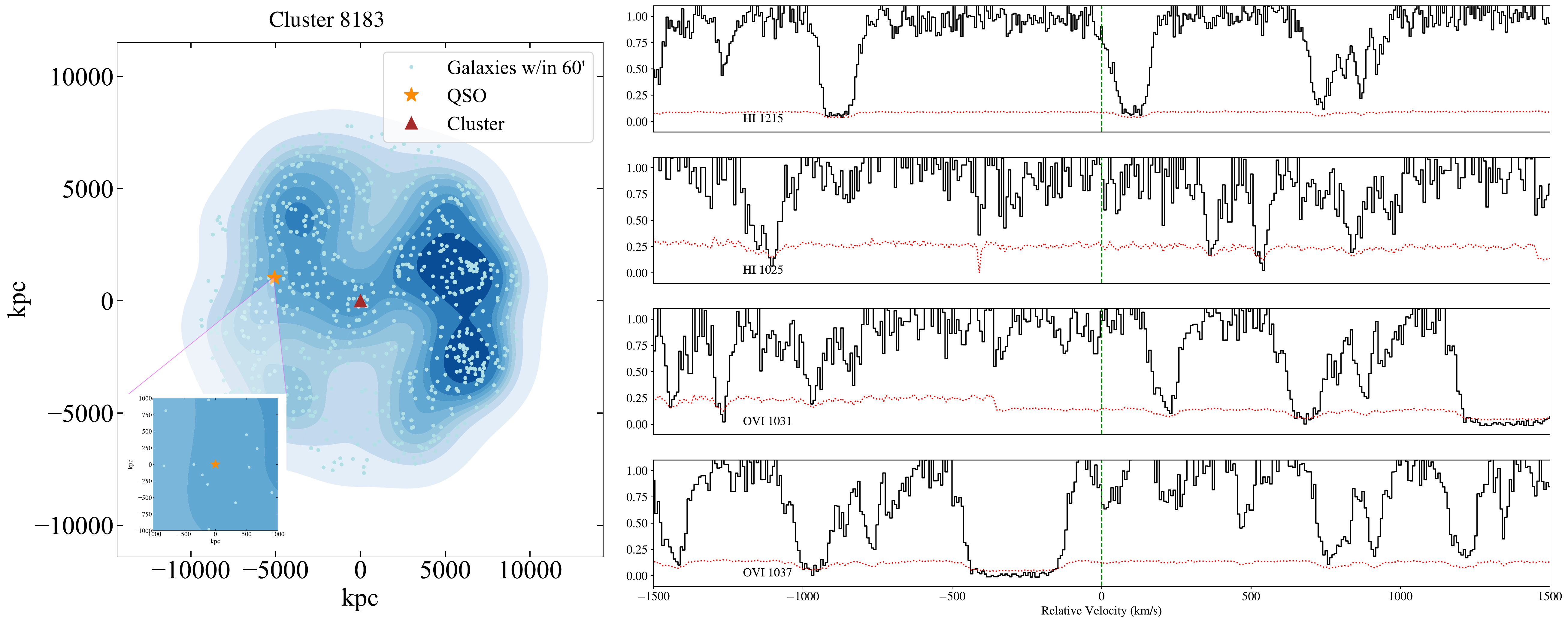}
\end{figure*}
\begin{figure*}
  \centering
  \includegraphics[width=\textwidth]{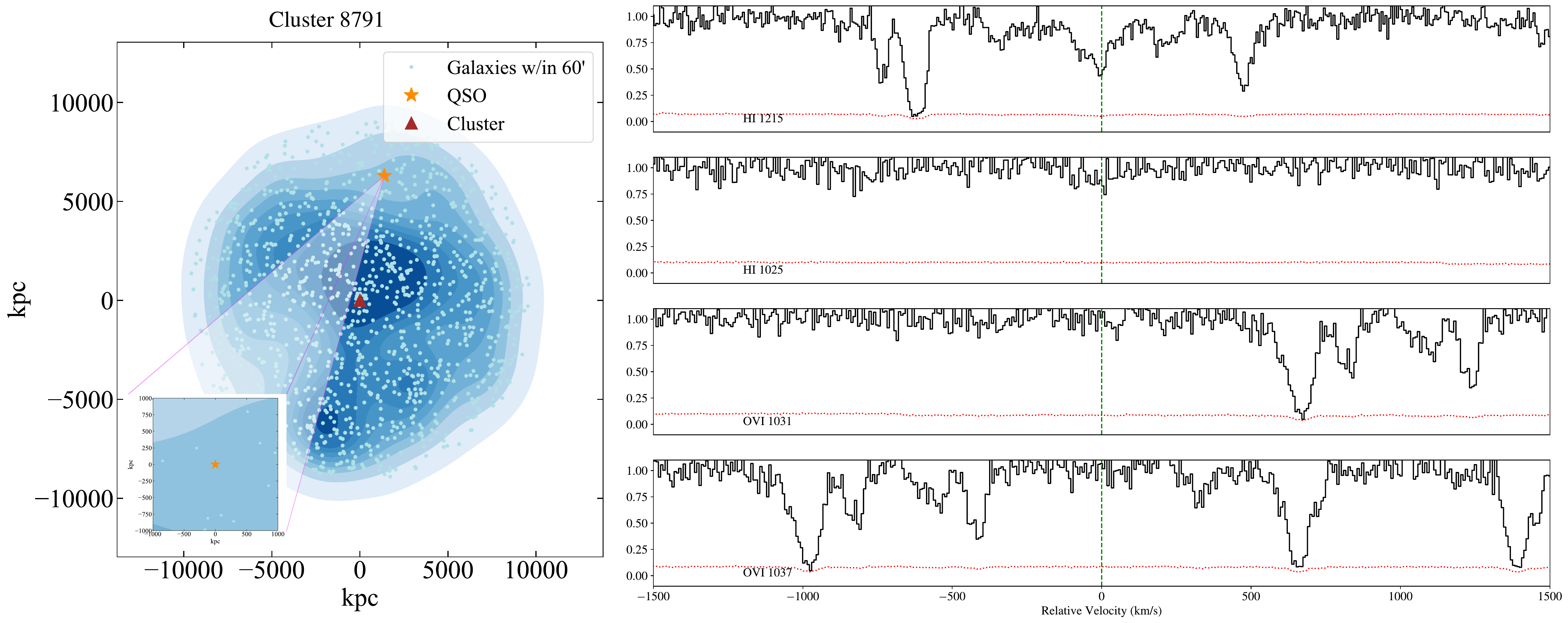}
\end{figure*}
\begin{figure*}
  \centering
  \includegraphics[width=\textwidth]{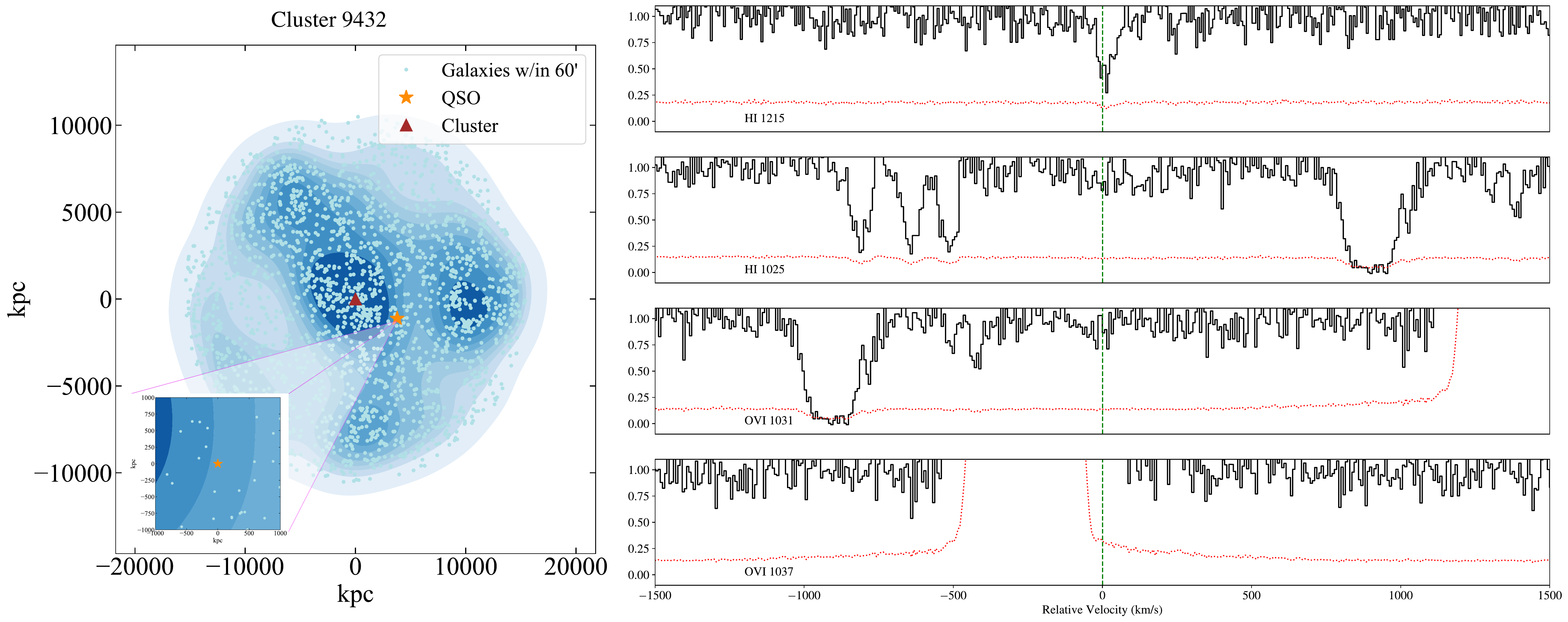}
\end{figure*}
\begin{figure*}
  \centering
  \includegraphics[width=\textwidth]{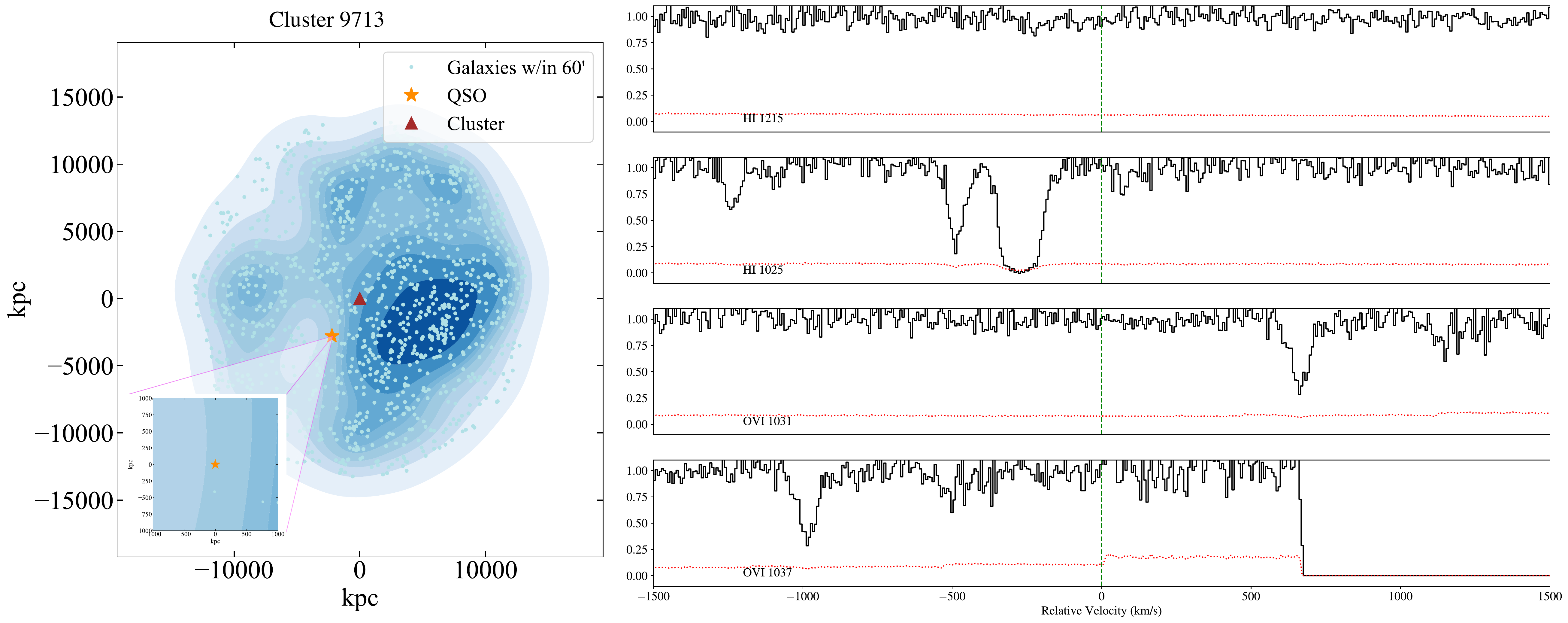}
\end{figure*}
\begin{figure*}
  \centering
  \includegraphics[width=\textwidth]{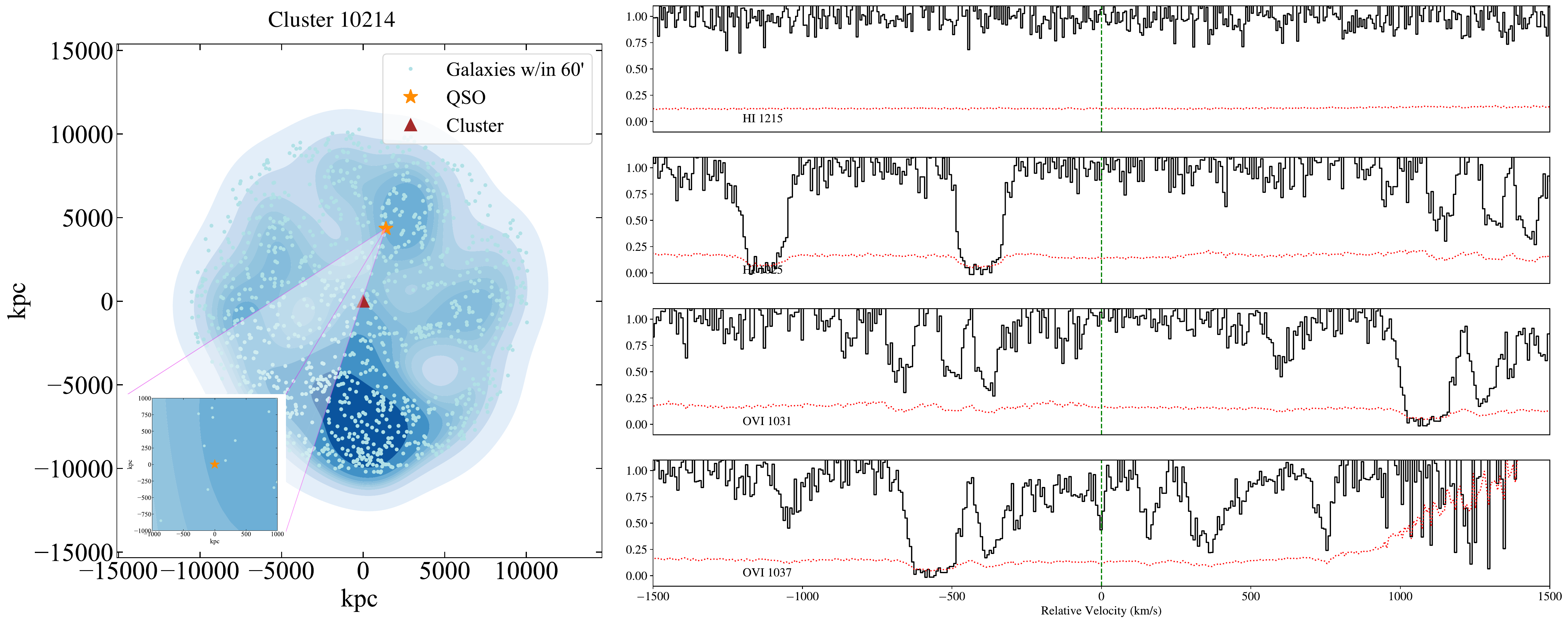}
\end{figure*}
\begin{figure*}
  \centering
  \includegraphics[width=\textwidth]{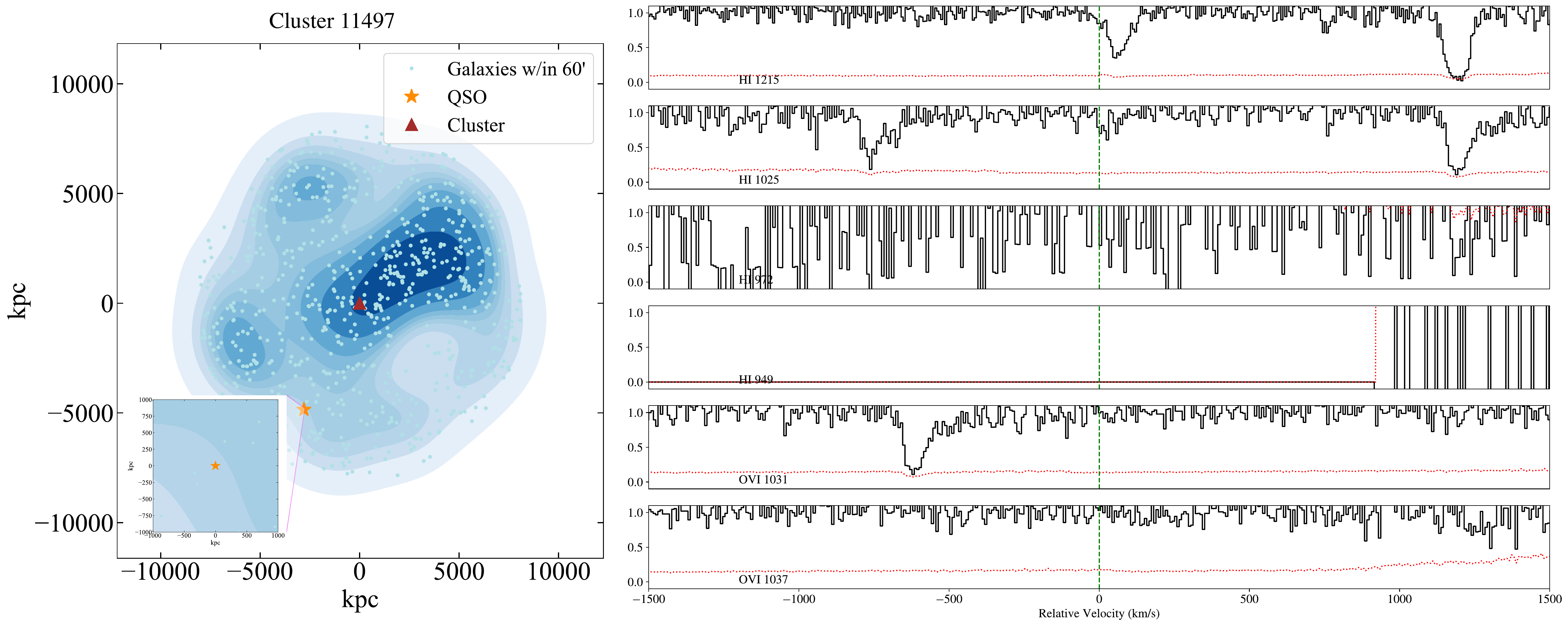}
\end{figure*}
\begin{figure*}
  \centering
  \includegraphics[width=\textwidth]{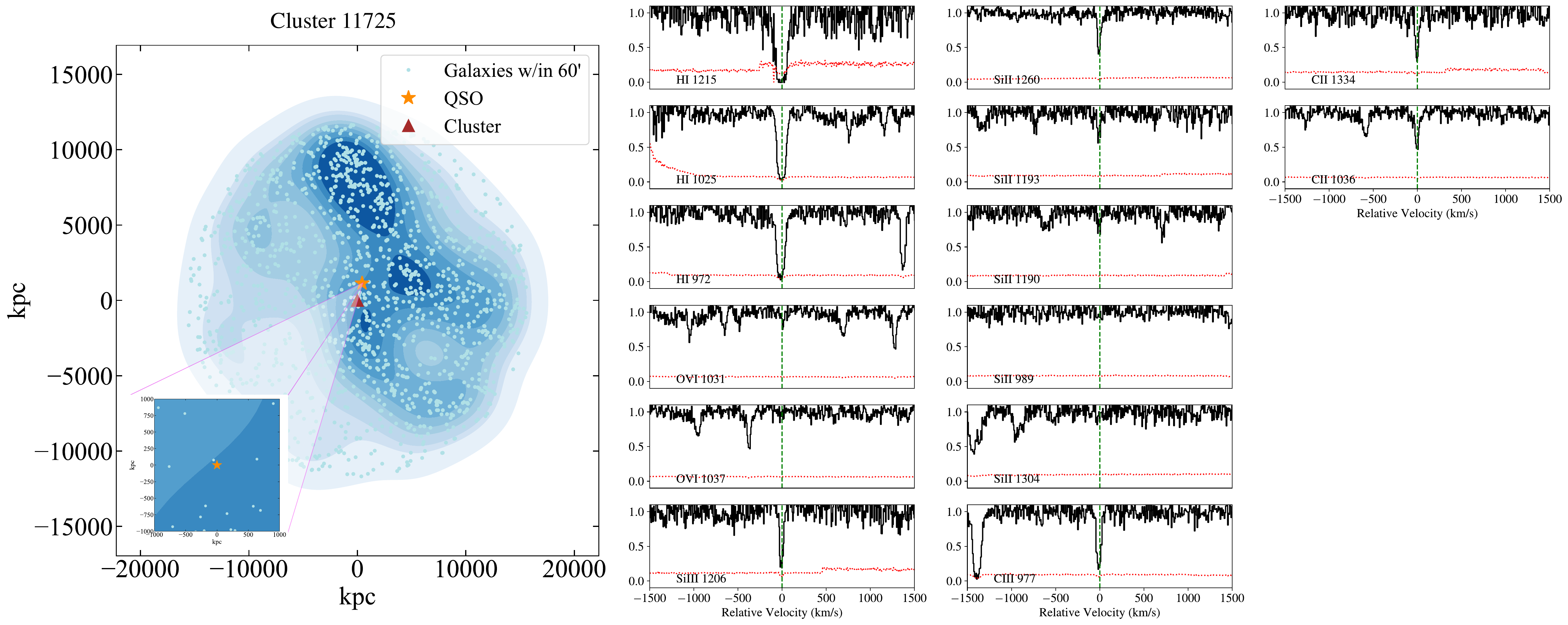}
\end{figure*}
\begin{figure*}
  \centering
  \includegraphics[width=\textwidth]{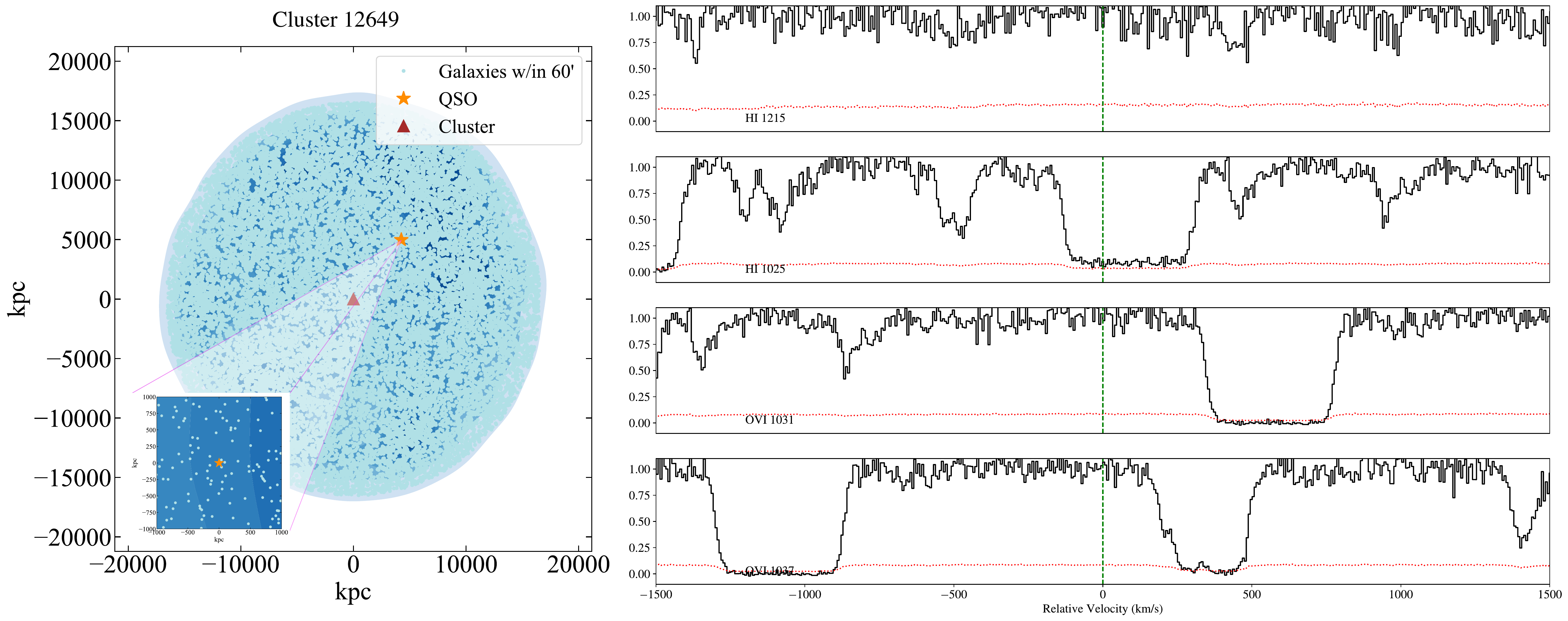}
\end{figure*}
\begin{figure*}
  \centering
  \includegraphics[width=\textwidth]{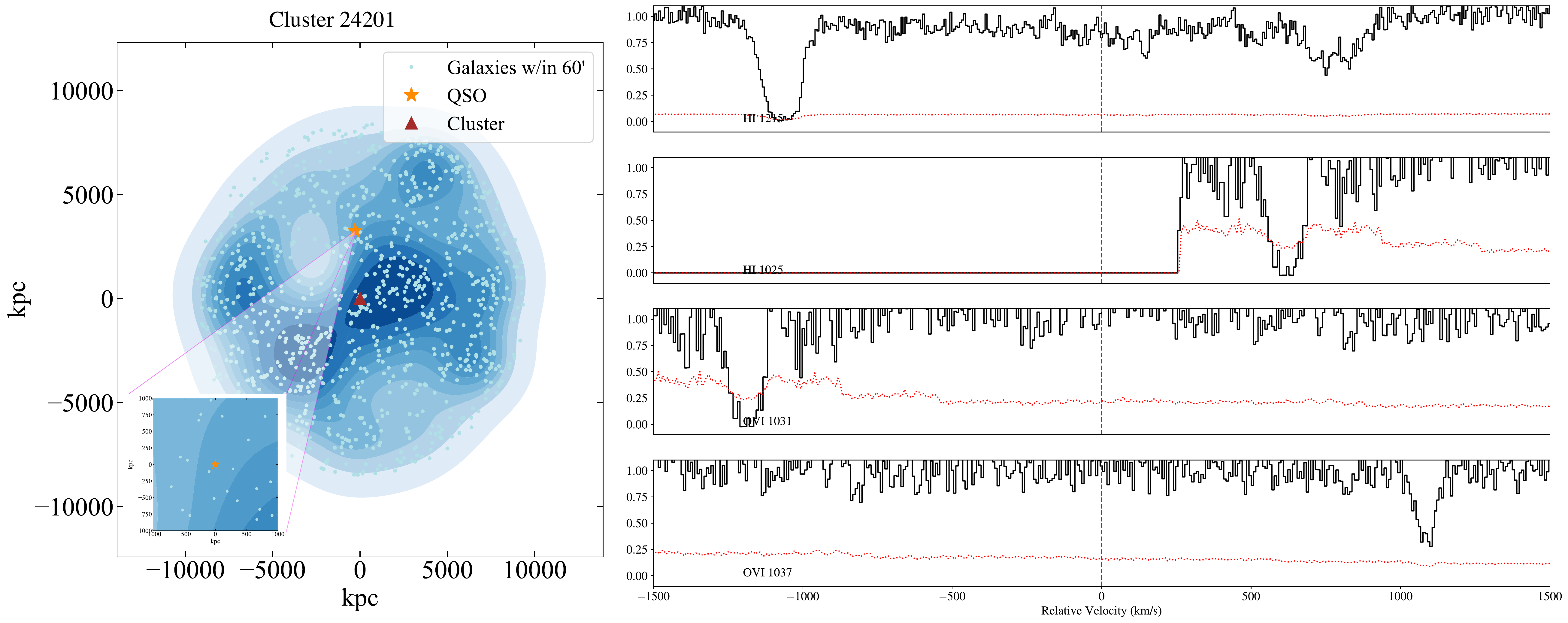}
\end{figure*}
\begin{figure*}
  \centering
  \includegraphics[width=\textwidth]{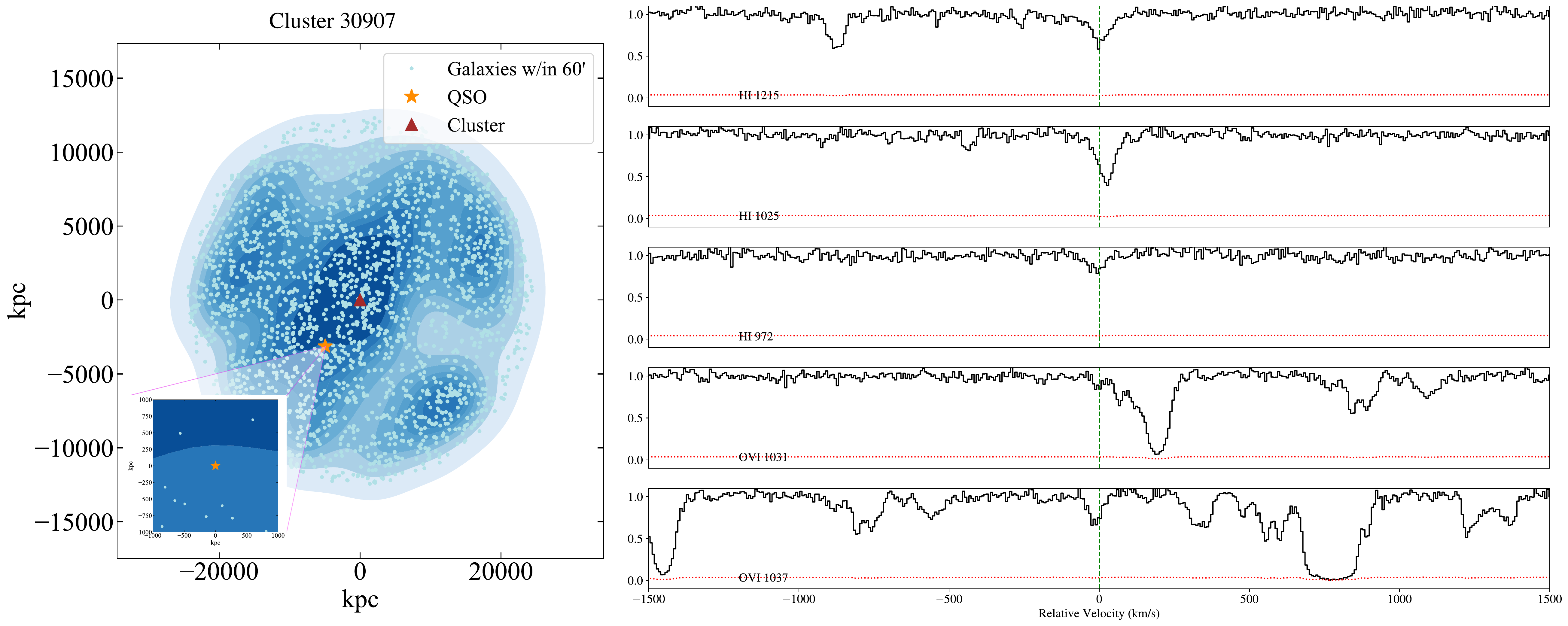}
\end{figure*}
\begin{figure*}
  \centering
  \includegraphics[width=\textwidth]{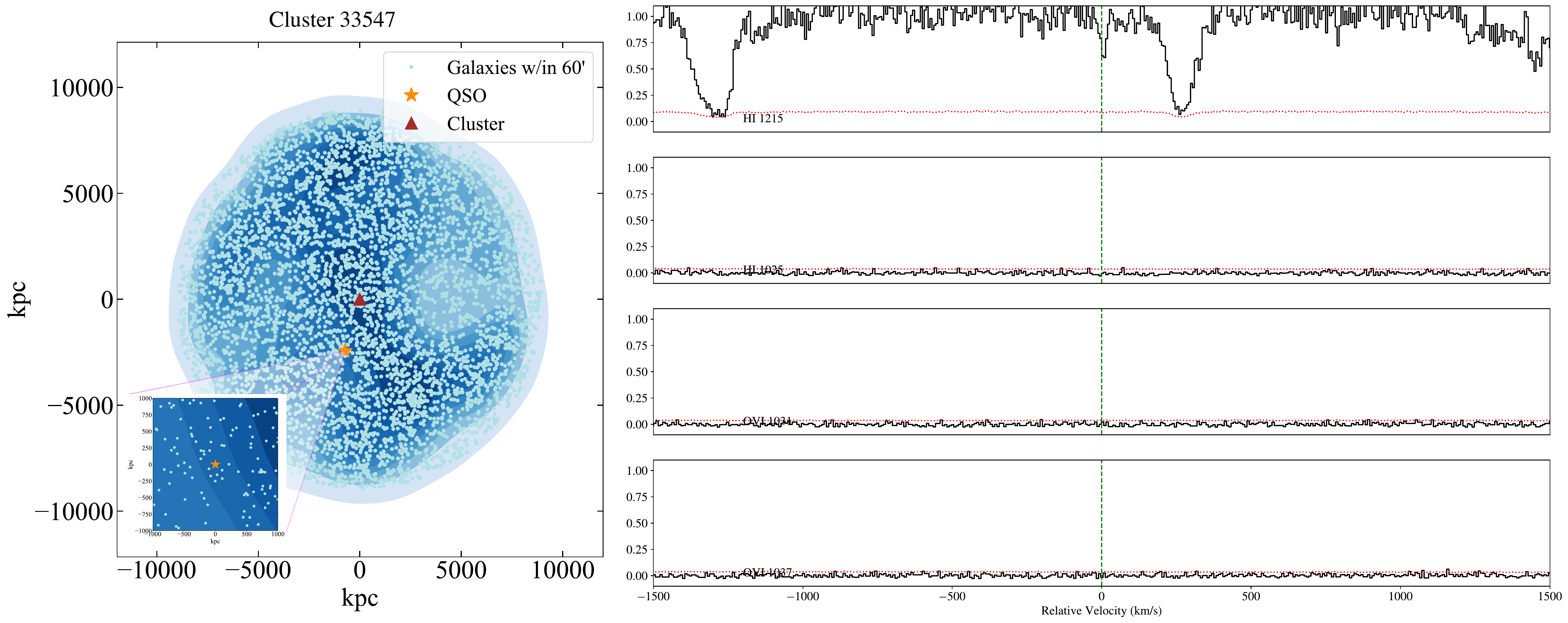}
\end{figure*}
\begin{figure*}
  \centering
  \includegraphics[width=\textwidth]{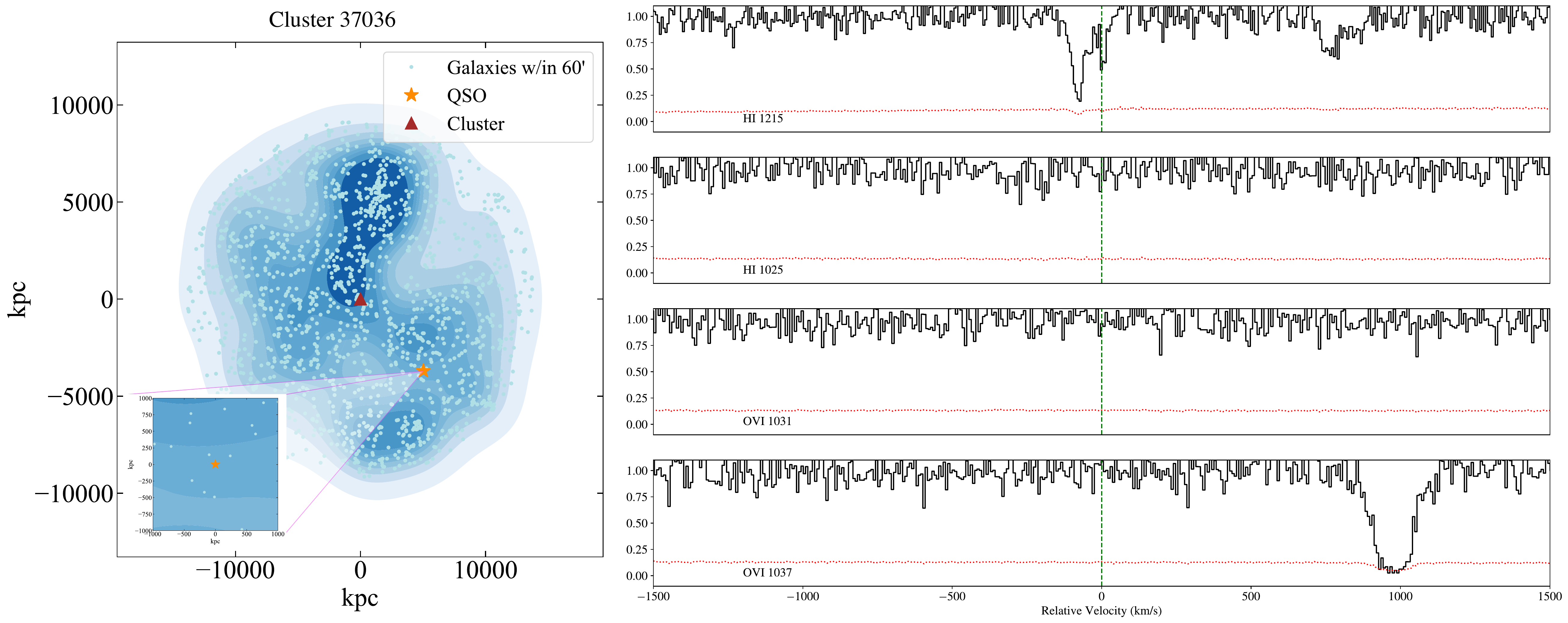}
\end{figure*}
\begin{figure*}
  \centering
  \includegraphics[width=\textwidth]{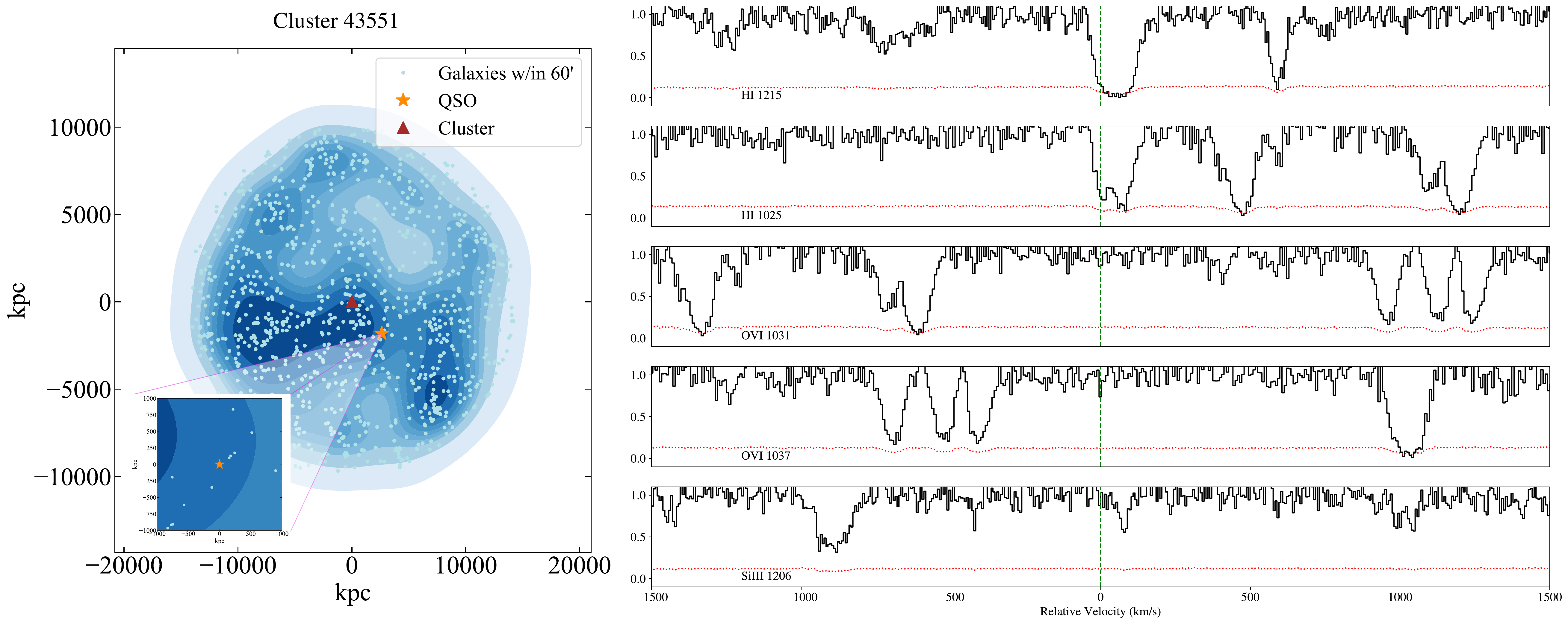}
\end{figure*}
\begin{figure*}
  \centering
  \includegraphics[width=\textwidth]{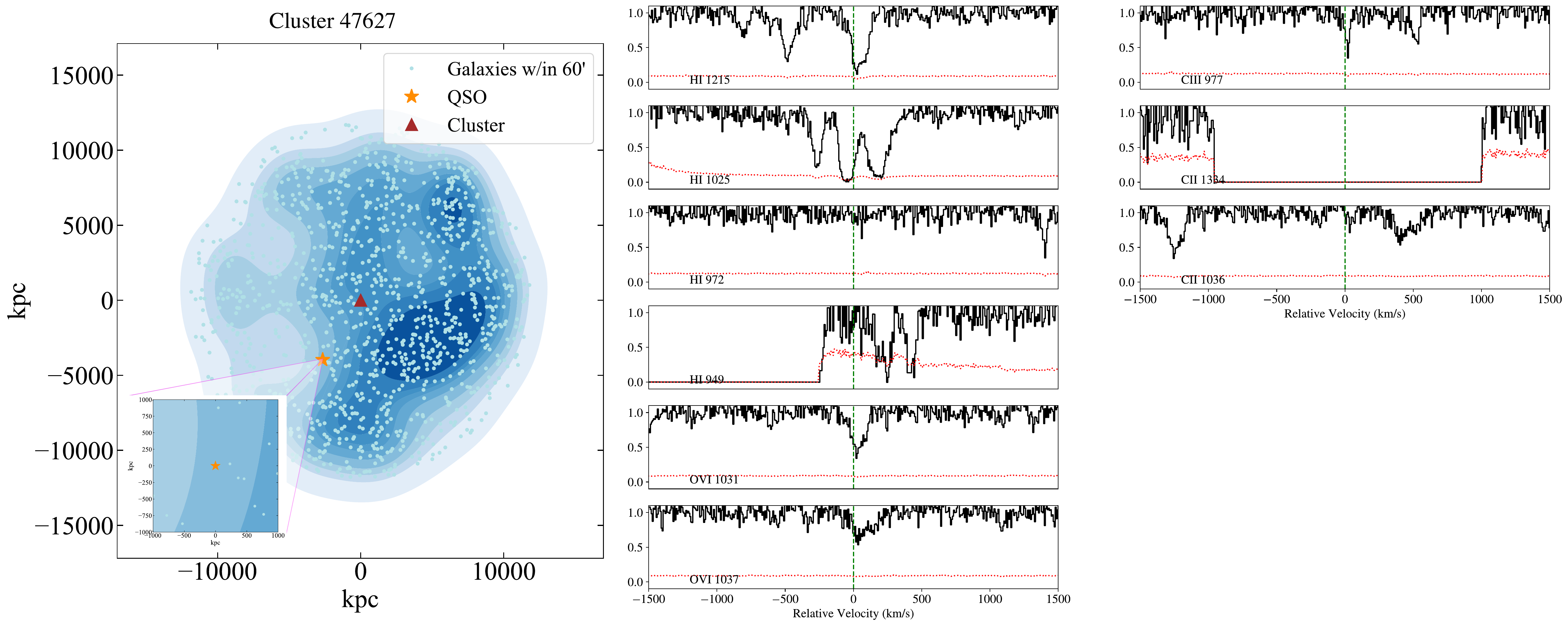}
\end{figure*}
\begin{figure*}
  \centering
  \includegraphics[width=\textwidth]{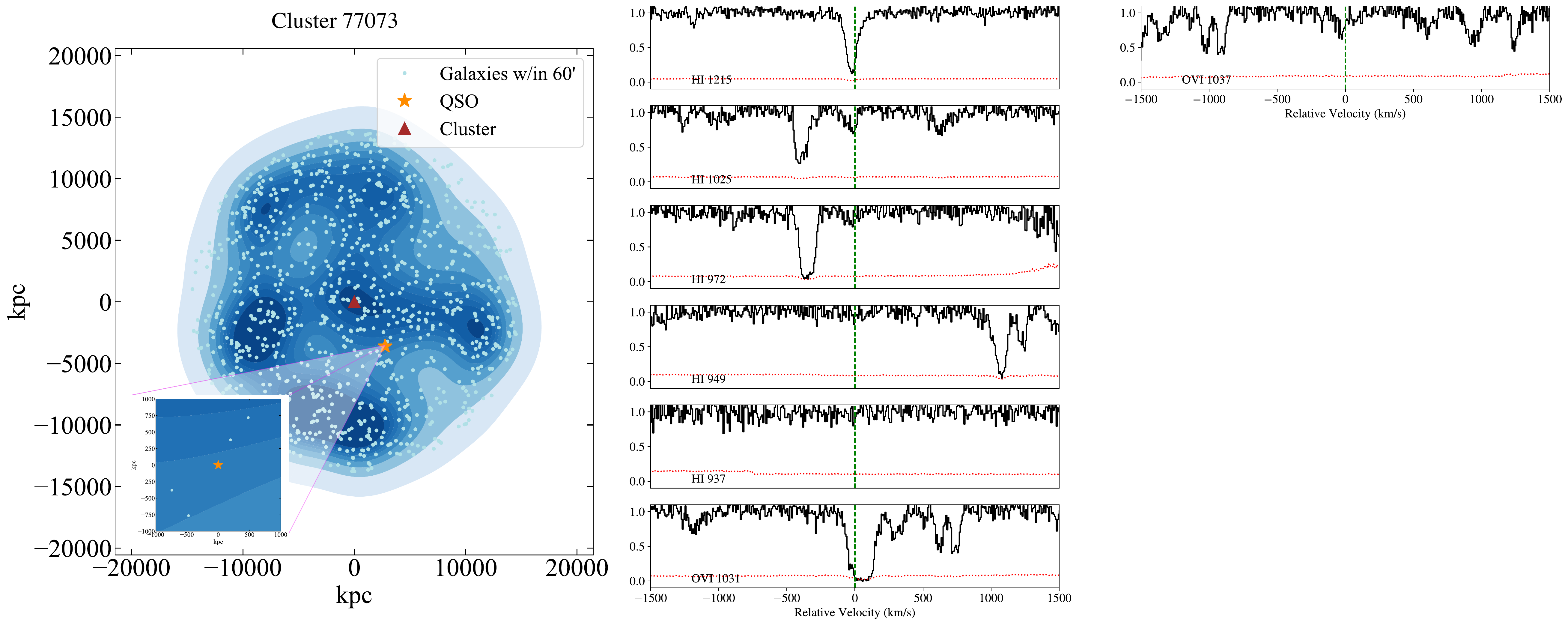}
\end{figure*}
\begin{figure*}
  \centering
  \includegraphics[width=\textwidth]{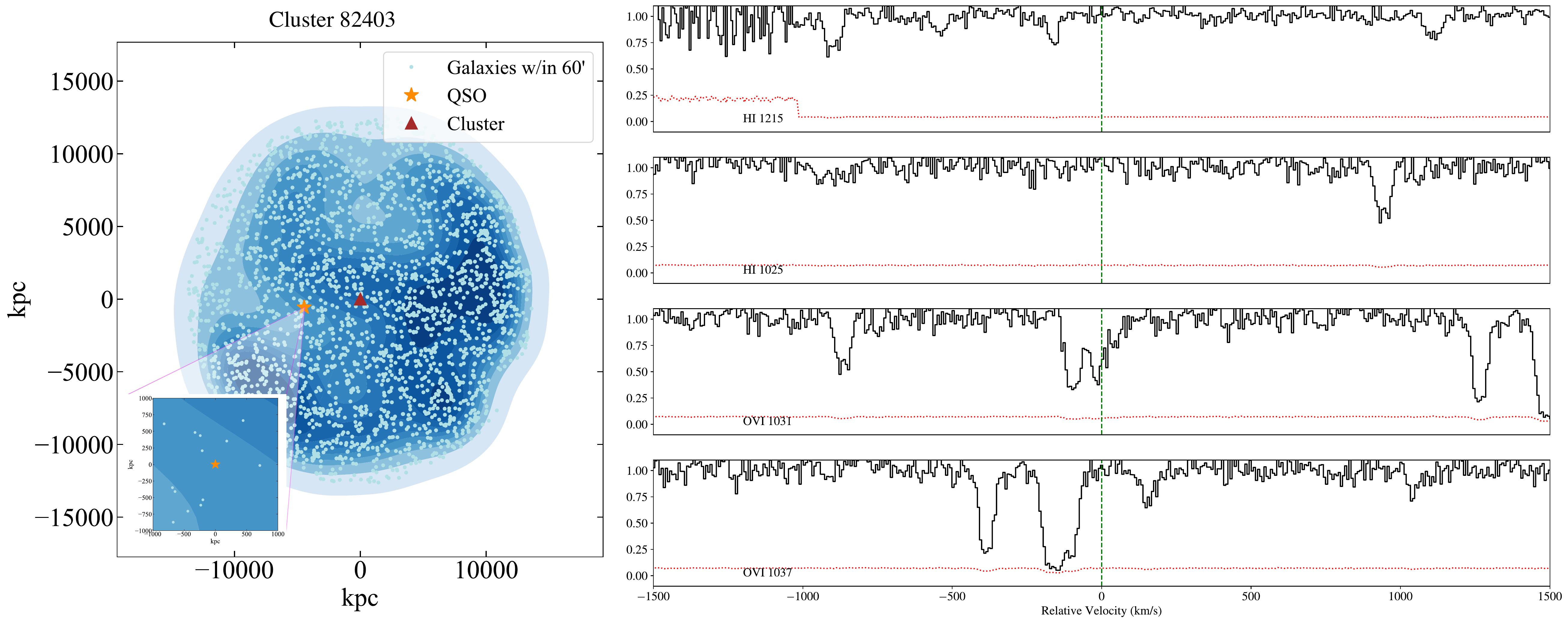}
\end{figure*}
\begin{figure*}
  \centering
  \includegraphics[width=\textwidth]{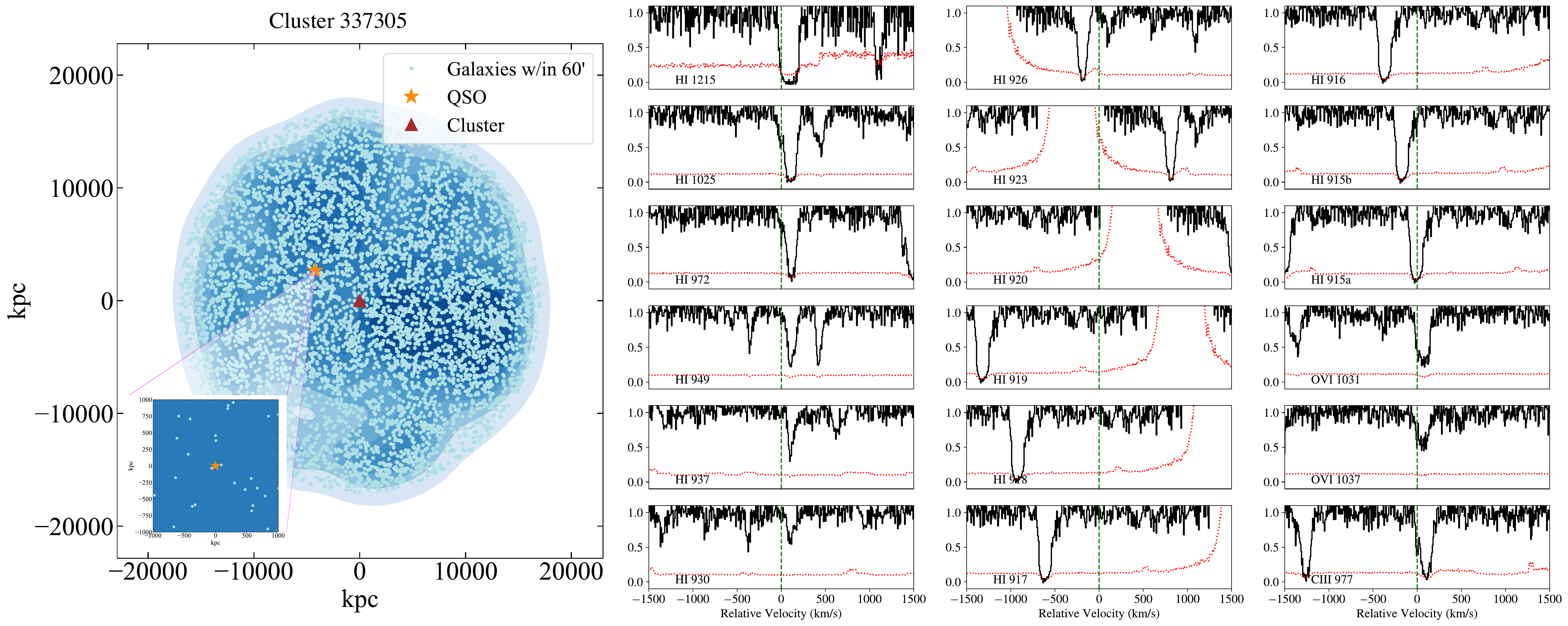}
\end{figure*}

%



\newpage
\pagebreak
\bibliography{refs2}{}
\bibliographystyle{aasjournal}



\end{document}